%% file: main.tex
\documentclass[sigconf]{acmart}
\AtBeginDocument{%
  \providecommand\BibTeX{{%
    \normalfont B\kern-0.5em{\scshape i\kern-0.25em b}\kern-0.8em\TeX}}}

\setcopyright{acmcopyright}
\copyrightyear{2024}
\acmYear{2024}
\acmDOI{X}

\acmConference[ASIACCS 2024]{
19th ACM ASIA Conference on Computer and Communications Security}{July 01--05, 2024}{Singapore}
\acmPrice{15.00}
\acmISBN{978-1-4503-XXXX-X/18/06}

\usepackage{tikz}
\usepackage{amsmath}

\usepackage{filecontents}

\usepackage{subcaption}
\usepackage{algorithm}
\usepackage{listings}
\usepackage{gensymb}
\usepackage[utf8]{inputenc}
\DeclareUnicodeCharacter{2212}{-}
\usepackage{mathtools}
\usepackage{algpseudocode}
\usepackage{multirow}
\usepackage{url}
\usepackage{hyperref}
\usepackage{xcolor,colortbl}
\usepackage{caption}    
\usepackage{booktabs}
\usepackage{pifont}
\usepackage{verbatim}
\usepackage{tikz}

\usepackage{makecell}

\usepackage{amsmath, amsfonts}
\usepackage{enumitem}
\usepackage{wrapfig}
\usepackage[htt]{hyphenat}

\usepackage{listings}
\usepackage{xcolor}
\usepackage{graphicx}

\definecolor{darkpastelgreen}{rgb}{0.01, 0.75, 0.24}
\definecolor{darkpastelred}{rgb}{0.76, 0.23, 0.13}
\usepackage{pifont}

\renewcommand{\paragraph}[1]{\medskip \noindent {\bf #1.}}

\lstdefinestyle{customc}{
  belowcaptionskip=1\baselineskip,
  moredelim=**[is][\color{red}]{@}{@},
  breaklines=true,
  frame=L,
  numbers=left,
  xleftmargin=13pt,
  numbersep=5pt, 
  language=C++,
  showstringspaces=false,
  basicstyle=\footnotesize\ttfamily,
  keywordstyle=[1]\bfseries\color{blue!40!black},
  commentstyle=\itshape\color{purple!40!black},
  identifierstyle=\color{black},
  stringstyle=\color{orange},  
  keywords=[2]{matched, authenticated, result, ret, fast\_auth\_result},
  keywordstyle=[2]\bfseries\color{darkpastelred},
  morekeywords={munmap},
}
\begin{document}

\title{Mayhem: Targeted Corruption of Register and Stack Variables}

\author{Andrew J. Adiletta}
\authornote{Both authors contributed equally to this research.}
\email{ajadiletta@wpi.edu}
\affiliation{%
  \institution{Worcester Polytechnic Institute}
  \city{Worcester}
  \state{MA}
  \country{USA}
}
\author{M. Caner Tol}
\authornotemark[1]
\email{mtol@wpi.edu}
\affiliation{%
  \institution{Worcester Polytechnic Institute}
  \city{Worcester}
  \state{MA}
  \country{USA}
}
\author{Yarkın Doröz}
\email{ydoroz@wpi.edu}
\affiliation{%
  \institution{Worcester Polytechnic Institute}
  \city{Worcester}
  \state{MA}
  \country{USA}
}
\author{Berk Sunar}
\email{sunar@wpi.edu}
\affiliation{%
  \institution{Worcester Polytechnic Institute}
  \city{Worcester}
  \state{MA}
  \country{USA}
}

\renewcommand{\shortauthors}{Adiletta et al.}


\input{sections/abstract}

\ccsdesc[500]{Stack Rowhammer Attack~Using Rowhammer to inject faults into stack and register variables}



\keywords{Rowhammer, Stack, Register Flipping}


\maketitle

\input{sections/introduction}

\input{sections/background}

\input{sections/stack_rowhammer}
\input{sections/experiment_results}


\input{sections/tls}
\input{sections/countermeasures}
\section{Acknowledgements} 
We thank our anonymous reviewers for their insightful feedback and Saleh K. Monfared for useful discussions on Linux. This work was supported by the National Science Foundation grant CNS-2026913 and in part by a grant from the Qatar National Research Fund.
\bibliography{references}
\appendix
\input{sections/appendix}



\newpage
\pagebreak

\bibliographystyle{plain}

\end{document}

%% file: sections/abstract.tex
\begin{abstract}

In the past decade, many vulnerabilities were discovered in microarchitectures which yielded attack vectors and motivated the study of countermeasures. Further, architectural and physical imperfections in DRAMs led to the discovery of Rowhammer attacks which give an adversary power to introduce bit flips in a victim's memory space. Numerous studies analyzed Rowhammer and proposed techniques to prevent it altogether or to mitigate its effects.

In this work, we push the boundary and show how Rowhammer can be further exploited to inject faults into stack variables and even register values in a victim's process. We achieve this by targeting the register value that is stored in the process's stack, which subsequently is flushed out into the memory, where it becomes vulnerable to Rowhammer. When the faulty value is restored into the register, it will end up used in subsequent iterations. The register value can be stored in the stack via latent function calls in the source or by actively triggering signal handlers. We demonstrate the power of the findings by applying the techniques to bypass SUDO and SSH authentication. We further outline how MySQL and other cryptographic libraries can be targeted with the new attack vector. There are a number of challenges this work overcomes with extensive experimentation before coming together to yield an end-to-end attack on an OpenSSL digital signature: achieving co-location with stack and register variables, with synchronization provided via a blocking window. We show that stack and registers are no longer safe from the Rowhammer attack.  
\end{abstract}

%% file: sections/introduction.tex
\section{Introduction}
The emergence of attacks such as Meltdown \cite{Lipp2018meltdown} and Spectre \cite{Kocher2018spectre} exposed intrinsic vulnerabilities in our computing infrastructure. These microarchitectural leakages were further developed and exploited in a number of studies, e.g.,  \cite{gras2017aslr,islam2019spoiler,vanschaik2019ridl,canella2019fallout,vanbulck2020lvi}.

\smallskip
\noindent
{\bf Rowhammer Fault Injection}
While these vulnerabilities focused on passive leakages, Rowhammer emerged as a realistic tool for an attacker to actively inject faults in DRAMs~\cite{kim2014flipping,seaborn2015exploiting}. Rowhammer exploits the fact that if a row in DRAM is accessed repeatedly, this may lead to bit flips in neighboring rows. Rowhammer has proven effective in real-life attack scenarios. For instance, ~\cite{gruss2018another} showed that it is possible to gain root access with opcode flipping in the \texttt{sudo} binary, \cite{flipfengshui} demonstrated an end-to-end attack breaking \texttt{OpenSSH} public-key authentication, \cite{weissman2019jackhammer} demonstrated a Bellcore attack on a CRT-based RSA implementation in \texttt{WolfSSL} to recover secret keys. Further pushing the boundaries, \cite{2016Rowhammerjs} and \cite{desmash} have shown that Rowhammer can be applied even remotely through JavaScript. Similarly, \cite{tatar2018thRowhammer} and \cite{lipp2020nethammer} demonstrated that Rowhammer can be executed over the network. Rowhammer is also applicable in cloud environments \cite{xiao2016one, cojocar2020we} and heterogeneous FPGA-CPU platforms \cite{weissman2019jackhammer}. 
Beyond DRAMs, \cite{Cai2017VulnerabilitiesIM} has shown that \textit{flash} memories are also prone to Rowhammer-like cell-to-cell interference, which then, when used to target file-system pages, can result in privilege escalation \cite{Kurmus2017FromRB}.

\paragraph{Rowhammer Countermeasures}
The severity of the threat motivated numerous Rowhammer countermeasures, e.g., for detection  \cite{irazoqui2016mascat,chiappetta2016real, zhang2016cloudradar, herath2015these, payer2016hexpads, gruss2016flush+, aweke2016anvil, corbet} and mitigation \cite{2016Rowhammerjs, van2016drammer, brasser2017can}. Unfortunately, \cite{gruss2018another} has shown that all of these countermeasures are ineffective. Further, \cite{cojocar2019ecc} showed that ECC, a hardware-enabled error checking built into many memory devices, can also be bypassed. Another hardware countermeasure Target Row Refresh (TRR), has also been recently defeated \cite{frigo2020trrespass}. This work was extended in \cite{desmash}, claiming that more than 80\% of the DRAM chips in the market are still vulnerable to Rowhammer. Quite recently, hammering beyond adjacent locations, i.e., HalfDouble~\cite{kogler2022halfdouble} hammering, was shown  to be effective in circumventing TRR mitigations. 


\paragraph{Faulting CPU Internals}
With significant efforts put into advancing Rowhammer attacks and countermeasures, one constant has been the assumption that CPU internals is impervious to software-based fault injection attacks. Specifically, SRAM-based registers and caches are assumed to be free from fault injection (except via direct physical manipulations such as in laser fault injection attacks). On the other hand, CPU-external devices such as DRAMs are greatly vulnerable to physical tampering. This view has been around since the early times of Trusted Computing and was motivated further by the introduction of cold-boot attacks~\cite{Halderman2008LestWR}.

In this work, we demonstrate that {\em CPU internals such as register values are also vulnerable}. Until now, Rowhammer attacks were generally targeted at corrupting dynamically allocated memory~\cite{mutlu2019Rowhammer} or binaries stored on disk loaded into memory~\cite{seaborn2015exploiting}. Other works \cite{yim2016rowhammer} have mentioned that there are key registers like \texttt{EIP} and \texttt{ESP} that if corrupted can affect the control flow of an x86 program, but cannot be corrupted if the register values are stored in a processor core. Here we show that register values can be forced by an attacker to be saved to the stack and flushed out to memory, where they become vulnerable to Rowhammer fault injection. Upon reload, the faulty values are reloaded into the registers before resuming execution. 


\paragraph{Targeting the Stack}
Besides flushed register values, vulnerable pieces of code exist within the stack of programs, e.g. security checks and authentication states. When these sensitive variables are corrupted, this may result in privilege escalation. Crypto libraries, for instance, minimize or eliminate dynamic memory and stack use either to support execution in constrained environments, or for safety-critical systems such as embedded or RTOS systems or to minimize exposure of potentially vulnerable internal secrets. The \texttt{wolfSSL} library, for instance, supports compilation options to avoid dynamic memory use. The NaCL library, in contrast, avoids dynamic memory and variable-size stack allocation altogether. Crypto library implementations, therefore, heavily rely on registers, and stack variables. Here we show that such variables are not secure against fault injection. Hence the attack surface of Rowhammer is greater than previously assumed.

\subsection*{Our Contribution}
In this paper, we systematically analyze the threat imposed by Rowhammer fault injections to stack variables and register values that were previously considered secure against Rowhammer. Specifically, we
\begin{itemize}[noitemsep,topsep=0pt,leftmargin=*]
    \item Introduce a novel attack to inject faults into register values through the stack memory;
    \item Show how static code/data allocation can be manipulated with bait pages to achieve co-location with the victim's stack;
    \item Introduce new synchronization techniques to enable practical means to target stack and register via Rowhammer; 
    \item Demonstrate attacks on \texttt{SUDO}, \texttt{OpenSSH}, and \texttt{MySQL};
    \item Highlight new RSA Bellcore vulnerabilities enabled by the attack vectors discovered in \texttt{OpenSSL};
    \item Demonstrate a full attack on code using OpenSSL for signature verification with attacks on both stack and register variables\footnote{
The source code is available on \url{https://github.com/vernamlab/mayhem}.
}
    \item Outline mitigative coding styles to minimize the attack surface against the newly introduced attack vectors.
\end{itemize}

\subsection*{Outline of the Paper}
The rest of the paper is organized as follows. Section~\ref{sec:background} gives background on our attack. In Section~\ref{sec:threat_model}, we explain the threat model of our attack. In Section~\ref{sec:flipping_bits_in_stack}, we explain the offline and online stages of the attack which include getting physically continuous memory and profiling bait pages. In Section~\ref{sec:flip_regs}, we explain injecting faults in stack memory and explain how to flip CPU register values using Rowhammer. In Section~\ref{sec:experiments}, we give the experimental evaluation. In Section~\ref{sec:attacks}, we explain our findings and results on \texttt{OpenSSL}, \texttt{sudo}, and \texttt{OpenSSH}. In Section~\ref{sec:vuln_analysis}, we give our analysis on RSA Bellcore Attacks on \texttt{OpenSSL} and \texttt{MySQL}. In section \ref{sec:full_attack} we demonstrate a full end-to-end attack on an OpenSSL client/server signature verification. In Section~\ref{sec:countermeasures}, we propose several countermeasures against our attack.

%% file: sections/background.tex
\section{Background}\label{sec:background}
\subsection{Rowhammer Attacks}

With increasing DRAM densities the chance for bit flips and reliability failures is increasing.
Hence, to retain data every DRAM row has to be continuously refreshed usually with 64 msec intervals.
Although refreshing the rows periodically helps preventing the data corruption, Kim et al. ~\cite{kim2014flipping} showed that frequent access to the neighbor rows causes faster charge leakage, which effectively causes bit flips before the next refresh. This is known as the Rowhammer effect~\cite{kim2014flipping}. Seaborn {\em et al.} \cite{seaborn2015exploiting} introduced the double-sided Rowhammer flipping the victim cells even faster.

Gruss {\em et al.} \cite{gruss2018another} introduced one-location hammering and achieved root access with opcode flipping in \texttt{sudo} binary in 2018. Gruss {\em et al.} \cite{2016Rowhammerjs} and Ridder {\em et al.} \cite{desmash} have shown that Rowhammer can be applied through JavaScript remotely. Tatar {\em et al.} \cite{tatar2018thRowhammer} and Lipp {\em et al.} \cite{lipp2020nethammer} have proved that it can be executed over the network. Rowhammer is also applicable in cloud environments \cite{xiao2016one, cojocar2020we} and heterogeneous FPGA-CPU platforms \cite{weissman2019jackhammer}. In 2020, Kwong {\em et al.} \cite{kwong2020rambleed} demonstrated that Rowhammer is not just an integrity problem but also a confidentiality problem. 

There have been many efforts on Rowhammer detections \cite{irazoqui2016mascat,chiappetta2016real, zhang2016cloudradar, herath2015these, payer2016hexpads, gruss2016flush+, aweke2016anvil, corbet} and neutralization \cite{2016Rowhammerjs, van2016drammer, brasser2017can}. Gruss {\em et al.} \cite{gruss2018another} have shown that all of these countermeasures are ineffective. Cojocar {\em et al.} \cite{cojocar2019ecc} in 2019 showed that the ECC countermeasure is not secure either. Another hardware countermeasure Target Row Refresh (TRR) has also been recently bypassed by Frigo {\em et al.} \cite{frigo2020trrespass}. This work was extended by Ridder {\em et al.} \cite{desmash} to attack TRR-enabled DDR4 chips from JavaScript and claim that more than 80\% of the DRAM chips in the market are still vulnerable to Rowhammer. Quite recently, hammering beyond adjacent locations was shown ~\cite{kogler2022halfdouble} to be effective in circumventing TRR mitigations. 

\subsection{Countermeasures in Crypto Libraries}
Physical fault injection attacks are well known among crypto practitioners \cite{Boneh2015OnTI}. Crypto libraries, especially ones designed for embedded platforms, have implemented countermeasures since the early 2000s. For instance, \texttt{OpenSSL} implements error checks in CRT-based exponentiation to thwart Bellcore attacks~\cite{Boneh2015OnTI}. Still, fault injection has proven effective in~\cite{DBLP:conf/eurosp/0002T19} to corrupt Elliptic Curve Parameters in the \texttt{OpenSSL} library. Further,
~\cite{mus2023jolt} demonstrated a Rowhammer fault injection vulnerability in WolfSSL that resulted in ECDSA key disclosure. The fault was injected during the signing operation with private ECC keys, which occur during a TLS handshake between client and server. \texttt{WolfSSL} addressed this vulnerability by implementing a series of checks during each stage of the signing process to detect if data has been tampered with, and \texttt{WOLFSSL\_CHECK\_SIG\_FAULTS} was released as a security measure\cite{nist_2022}. Importantly, these checks that protect dynamic memory operate on the idea that variables in the stack are safe from Rowhammer, which this paper will demonstrate is not the case.

%% file: sections/stack_rowhammer.tex
\section{Threat Model for Memory Mayhem}\label{sec:threat_model}
We will explain the attack scenarios in detail for each attack target in Section~\ref{sec:attacks}.
In line with the previous Rowhammer attacks~\cite{kim2014flipping,gruss2018another,xiao2016one,cojocar2020we,2016Rowhammerjs}, we assume attacker-victim co-location in the same system. Co-location is a common assumption for many micro-architectural side-channel attacks~\cite{Lipp2018meltdown,Kocher2018spectre,canella2019fallout,vanbulck2020lvi,vanschaik2019ridl}. 
We assume the operating system works as intended without any compromise in its integrity and the attacker has user privileges throughout the paper. We assume the attacker does not have access to any service that reveals the physical address or DRAM addressing information. Our attack does not require huge-page configuration and works with standard-size pages.  

\section{Flipping Bits in the Stack and Register Variables}\label{sec:flipping_bits_in_stack}
For the Rowhammer attack on DDR4 memory, we perform a multisided attack to circumvent TRR protection. We found that a multisided attack with 11 rows was most effective at getting flips on our system and we used \texttt{mfence} to prevent out-of-order execution. It is possible that without \texttt{mfence} CPU optimizations would disrupt the critical order that rows are accessed for the multisided attack which would prevent the attack from working. 
We found 1M accesses of all the rows were optimal in getting flips and reducing profiling/online time. We also found that doing 100 iterations of 1M accesses along with 100K \texttt{nops} in between also improved the efficacy of the attack in getting flips.

\subsection{Offline Memory Profiling}
\label{offline_mem_profiling}

Rowhammer requires that rows in DRAM are adjacent to each other physically. We achieve this through the use of the SPOILER and Row Conflict attacks. We use SPOILER ~\cite{islam2019spoiler} because it leaks virtual to physical address translation without the need to read the \texttt{pagemap} file, which would require root access. SPOILER takes advantage of a microarchitecture optimizations speculative loads and store forwarding. 
For finding addresses that are within the same bank, we use row conflicts ~\cite{pessl2016drama}, which is another timing side channel that we exploit to colocate memory for Rowhammer.

\paragraph{Profiling for Contiguous Memory} SPOILER first allocates a large buffer in the memory of the attacker program. The memory from this buffer is distributed throughout the DRAM randomly. Within a window of the memory buffer, SPOILER writes zeros to all the addresses, then times how long it takes to load the first entry in the array. 
Physical memory dependency requires more cycles to complete and thus would appear as peaks on the graph. For every system, the threshold values of SPOILER need to be adjusted. These threshold values include the timing required to call a memory read an outlier in the dataset (a timing measurement above a certain value is probably the result of a system interrupt or some other event rather than physical continuity), and a timing threshold value to qualify a value as a peak and thus part of the continuous memory buffer. 

In our experiments, we generally looked for about 3-5\% of our memory allocated to SPOILER to be physically continuous. This means that if we allocated 1024~MBytes of memory to our buffer, we would expect to find around 32-64 bytes of continuous memory. This varies depending on the experiment and the machine the experiment is running on. 

\paragraph{Finding Rows in the Same Bank} In addition to finding memory that is physically continuous, the memory must also be in the same bank of the DRAM for Rowhammer to work. We use the row conflict side channel to leak DRAM information which, like SPOILER, does not require root access. Rowconflict reads from the first address in the physically continuous memory buffer, then it reads from address $n$ (where $n$ is $1$ through the length of the memory buffer) and calculates the time difference between reads. A larger time difference indicates that the row buffer within the DRAM bank needed to be cleared and thus caused a spike in timing. Just like SPOILER, row conflict needs threshold values to be experimentally determined and defined for each machine. 

\subsection{Profiling for Bait Pages}
In order to flip a variable in the stack of a program, the page that the variable is located in needs to be placed in a page that has a flippable bit at the correct page offset. There are a number of pages used by the victim process that are irrelevant to our attack and would fill our flippy page before the page with our target variable. Thus, we must release unused filler pages we call \textit{bait pages}, which are filled with the victim's data that is irrelevant to the attack first. To flip a variable that is stored in a register and pushed to and popped from stack, a similar process is applied, but more complex profiling is necessary by manually looking at the memory space in the Linux kernel for the register value.   

\paragraph{Bait Page Profiling For Stack Attacks}The number of bait pages that need to be released depends on the process and if ASLR is enabled. The pseudo-code for releasing bait pages is given in Listing~\ref{lst:bait_pseudo}. We can see that the profiling process first allocates pages into its own process space to be released as bait for the victim process. It then unmaps the flippy page (this happens in the online stage) and unmaps the bait pages so they get filled first. 

During the offline stage, we determine the proper number of bait pages to release by first releasing a large number bait pages (500 or more) and recording all the physical addresses of the released pages into a text file. Then, we launch the victim process and translate the virtual address of the target variable into a physical address, which we then searched for in the text file with released addresses. The index of the physical address in the text file determined the number of bait pages we needed to release. Although there is certain variability in how many of the bait pages are consumed by the victim process before it allocates the target variable to a page, experimentally we found the victim process will consume the same number of bait pages 30\% of the time as stated in section \ref{sec:success_rating}. 

\paragraph{Bait Page Profiling For Register Attacks}
Registers also fall victim to the same bait pages attack, but profiling is more difficult because they do not have a virtual address that can be translated into a physical address which can be found in the bait pages released. Instead, during the profiling stage, we edit the victim process to give the register a unique value (like \texttt{0xDEADBEEF}), then look into the processes memory with \texttt{\textbackslash proc\textbackslash PID\textbackslash mem} and look of the unique value. This is the effective virtual address of the register when it gets pushed to stack, and will get put back into the stack when its popped off. We can use the same method of converting the virtual address to a physical address using the PID and the pagemap file for the process. Importantly, editing the source code to add a unique value to the register is only necessary during the offline stage. During the online stage, the source code for the victim remains untouched, and the number of bait pages consumed before the registers are pushed to stack remain the same. 

\begin{figure}[!h]
\begin{lstlisting}[frame=single,
                    language=C++,
                    label={lst:bait_pseudo},
                    caption= Pseudo code showing how pages can be forced into a specific area in memory using a mapping-unmapping technique ]
buffer = mmap(baitPages * PAGESIZE)
munmap(flippyPageAddr, PAGESIZE)
for(i = 0; i < bait_pages; i++)
    munmap(&buffer[i*PAGESIZE], PAGESIZE)

\end{lstlisting}
\end{figure}

 We release the bait pages before the flippy page, as seen in Listing \ref{lst:bait_pseudo}. This is because the Linux Buddy Allocator algorithm that is used to allocate memory to different processes effectively acts like a last-out-first-in system, where the latest pages released to memory are used first.

\subsection{Online Attack Phase}
\paragraph{Releasing The Flippy Page and Bait Pages}
The first stage of the attack is releasing the flippy page found during the offline profiling stage, then releasing the correct number of bait pages also found during the profiling stage. It is important to immediately launch the victim process once the bait pages are released and to start the process in the same way that it was started during the profiling stage.

\paragraph{Attacking Processes After Sending SIGSTOP}
In practice, the victim process cannot be altered to send a signal when it is ready to be attacked and wait for a signal that the attack has finished. Instead, we can use the \texttt{SIGSTOP} signal to stop the program's execution and create a probabilistic model to determine if the process has stopped in the correct place in the process execution to attack the variables. After the variables have been attacked, the \texttt{SIGCONT} signal can be sent to continue its execution. For an attacker to have permission to send a signal, it must belong to the same session. This is special to the \texttt{SIGCONT} signal \footnote{\url{https://www.sudo.ws/docs/man/1.8.10/sudo.man/}}. 

\paragraph{Attacking Processes During a Blocking Window} The most optimal scenario for an attack is for vulnerable code to have a blocking window where the process is waiting for an event that may be triggered by the attacker. For \texttt{SUDO}, this could be the period where the process is waiting for the attacker to enter a password. The process saves state data to stack while waiting for the user to submit a password. High-level examples of synchronizing blocking codes are Password Input, IP Socket Connections, Signal Interrupts, Media Uploads, Other User Input.

\begin{figure}
    \centering
    \includegraphics[width=0.7\columnwidth]{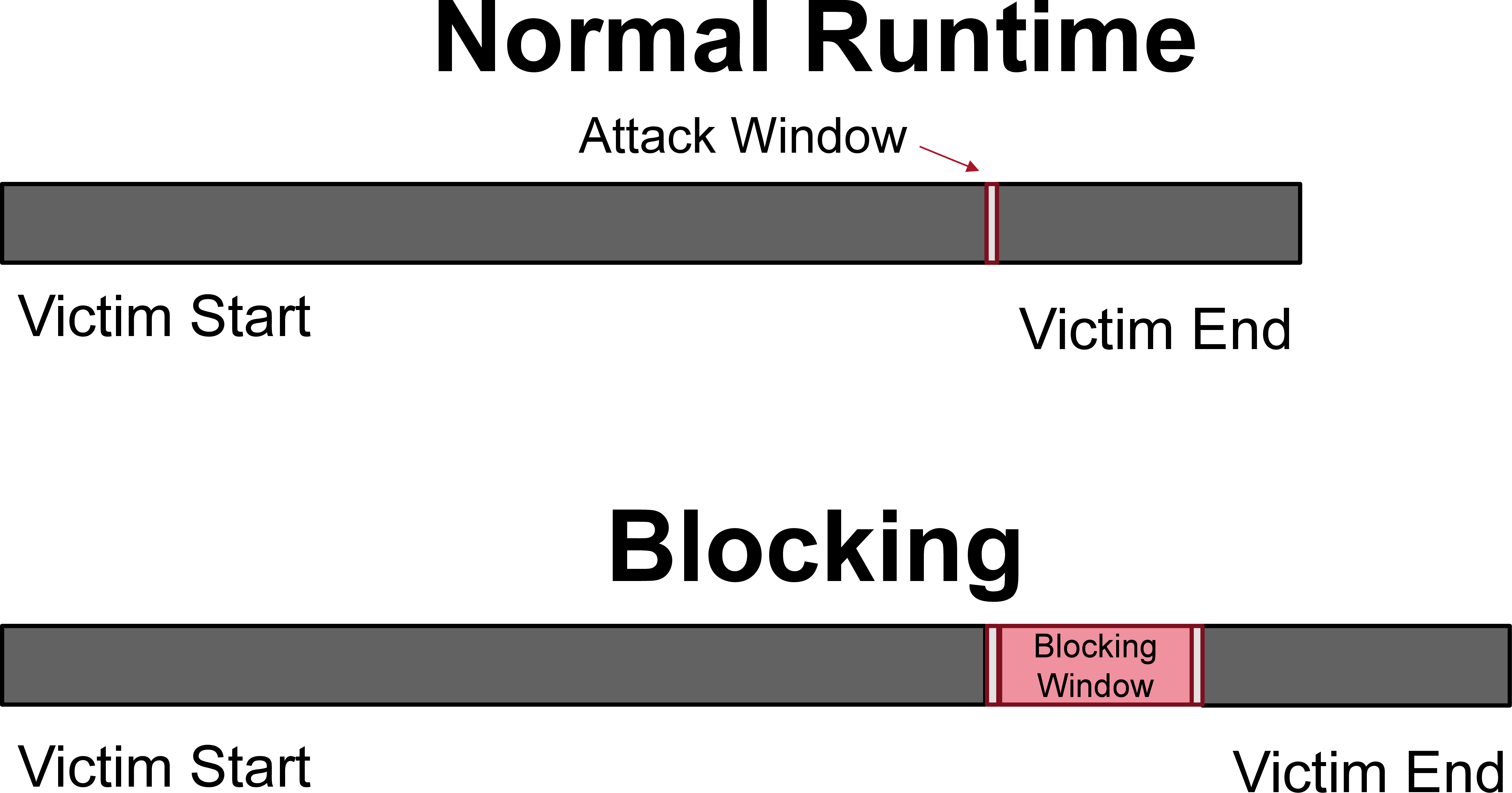}
    \caption{Diagram showing the run time of a program with a blocking window allowing the attacker to attack at the right time}
    \label{fig:blocking}
\end{figure}
\raggedbottom

Looking at Figure \ref{fig:blocking}, we can see that there is a blocking period during the attack window. This allows for synchronization between the victim and attacker so there is no longer a question of probability of the attacker will launch the Rowhammer attack at the right time. Practically, we can see an example of this on Section \ref{sec:IP_sync} of syncronization on a real-world TLS handshake. 

\subsection{Multisided Rowhammer}
There have been many efforts to detect Rowhammer by tracking consecutive reads to adjacent rows in a Serial Presence Detect (SPD) chip Intel CPUs deploy a mitigation known as pseudo-TRR or pTRR, which reads the Maximum Activation Count, or MAC value from the SPD and if reads to consecutive rows reaches the MAC value, the Intel CPU refreshes the row. 

A multisided attack works even with TRR enabled, with a strategy based on the Trrespass multisided attack \cite{frigo2020trrespass}.



\paragraph{Flipping Bits using Rowhammer}
The final step in the workflow after the target variable is loaded into memory is to actually flip the bits in the variable. While the profiling step allowed us to evaluate which bits were flipped in a row, because we do not control the area of memory being flipped, we cannot see which bits specifically were flipped. However, generally, the success of an attack can be determined by checking the new state of the process. For example, if the attack objective was achieved the attacker may bypass password authentication.

\section{Flipping Bits in CPU Registers}\label{sec:flip_regs}
The stack is a memory section that software processes use to store values temporarily. We experimentally found the context switching during moderate process activity and the attacker process running was enough to evict the victim data from the cache to be corrupted in DRAM without the attacker explicitly flushing it. 
The location of the last variable inserted into the stack is saved in the stack pointer registers. In assembly language, the stack can be used freely to store variables. However, higher-level languages and compilers use a convention that is based on the architecture and the operating system. These conventions set rules for converting C code into an assembly code, such as System V i386, System V x86\_64, Microsoft x64, and ARM. Each convention  uses different registers for function inputs and return variables, and in certain cases the convention also uses the stack to store temporary variables. This makes the variables vulnerable to fault attacks by using Rowhammer on the stack. Now we will summarize the convention to show which situations cause the compiler to use stack for variable storage. Since our setup is focused on Linux, we will focus on System V x86\_64. 

\subsection{Forcing Register Eviction to Stack} \paragraph{Intel-Ubuntu C convention}The architecture uses 16 64-bit registers which are referred to as \texttt{rax}, \texttt{rbx}, \texttt{rcx}, \texttt{rdx}, \texttt{rbp}, \texttt{rsp}, \texttt{rsi}, \texttt{rdi}, \texttt{r8-15}
. Some of the registers are special purpose registers, e.g. \texttt{rsp}: register stack pointer and others are generic/scratchpad registers. When a C code is compiled and converted into assembly code the following convention is used for functions: 

\begin{itemize}[noitemsep,topsep=0pt,leftmargin=*]
    \item \texttt{rax} holds the return value of the function
    \item \texttt{rdi}, \texttt{rsi}, \texttt{rdx}, \texttt{rcx}, \texttt{r8}, \texttt{r9} holds the input parameters of the function. If there are more than 6 input parameters, rest is written into the stack. 
    \item \texttt{rax}, \texttt{rdi}, \texttt{rsi}, \texttt{rdx}, \texttt{rcx}, \texttt{r8-11}
    are used as scratch registers. 
    \item \texttt{rax}, \texttt{rdi}, \texttt{rsi}, \texttt{rdx}, \texttt{rcx}, \texttt{r8-11}
    are caller-saved registers. This means that if a routine calls a subroutine, it is the responsibility of the main routine to preserve the values of any relevant registers, as the subroutine is free to modify them. To do this, the calling function can save these values in other registers that will not be changed during the subroutine call or save them on the stack.  
    \item \texttt{rbx}, \texttt{rsp}, \texttt{rbp}, \texttt{r12-15}
    are callee-saved registers. When a routine makes a subroutine call, it is the responsibility of the subroutine to ensure that the values of the relevant registers remain unchanged after the subroutine call is completed. To achieve this, the subroutine pushes the contents of these registers onto the stack and then restores the original values when it has finished executing by popping them from the stack.    
    \item When a function call has a large number of variable declarations, compilers aim to utilize as many registers as possible to store these values in order to optimize performance. However, when the number of available registers is insufficient to hold all the variables, the compilers will resort to using the stack to store the excess variables.
\end{itemize}

When the compilers use the Intel-Ubuntu C convention, the excess variables are stored on the stack if a function has many variables. This makes the variables vulnerable to stack attacks. Our inspection of disassembled code of common libraries shows that these cases are less common as compilers aim to reduce stack usage, but there is still a possibility. Of course, the attack can only be executed if the targeted variable is written to the stack. 
To enable the stack attack, we can force processes to temporarily store register contents on the stack during the execution of another process. This expands the scope of the attack beyond just variables stored on the stack. 

Below, we will discuss two methods to attack the stack variables. 


\paragraph{Passive} The first method exploits a natural occurrence. When a compiler uses the C convention to create executable code, it occasionally stores register values in the stack for safekeeping, e.g., \texttt{push} instruction is used at the beginning of the function calls. It is hard to mitigate since it is not visible in the source code, which makes most of the libraries potentially vulnerable depending on the compiler optimizations. Higher levels of optimization settings in compilers are more aggressive in using registers. 


\begin{wrapfigure}{l}{0.41\linewidth}
\vspace{-3mm}
\begin{lstlisting}[frame=single,
                    numbers=none,
                    breaklines=true, % Enables line breaking
                    label={lst:example_push},
                    caption=Example of push from LibC recv function]
<__recv@@GLIBC_PRIVATE>:
  ...
  push   %ebx
  ...
  %ebx,%eax
  pop    %ebx
  %esi
  ret  
\end{lstlisting}
\end{wrapfigure}
\begin{figure}
    \centering
    \includegraphics[width=0.8\columnwidth]{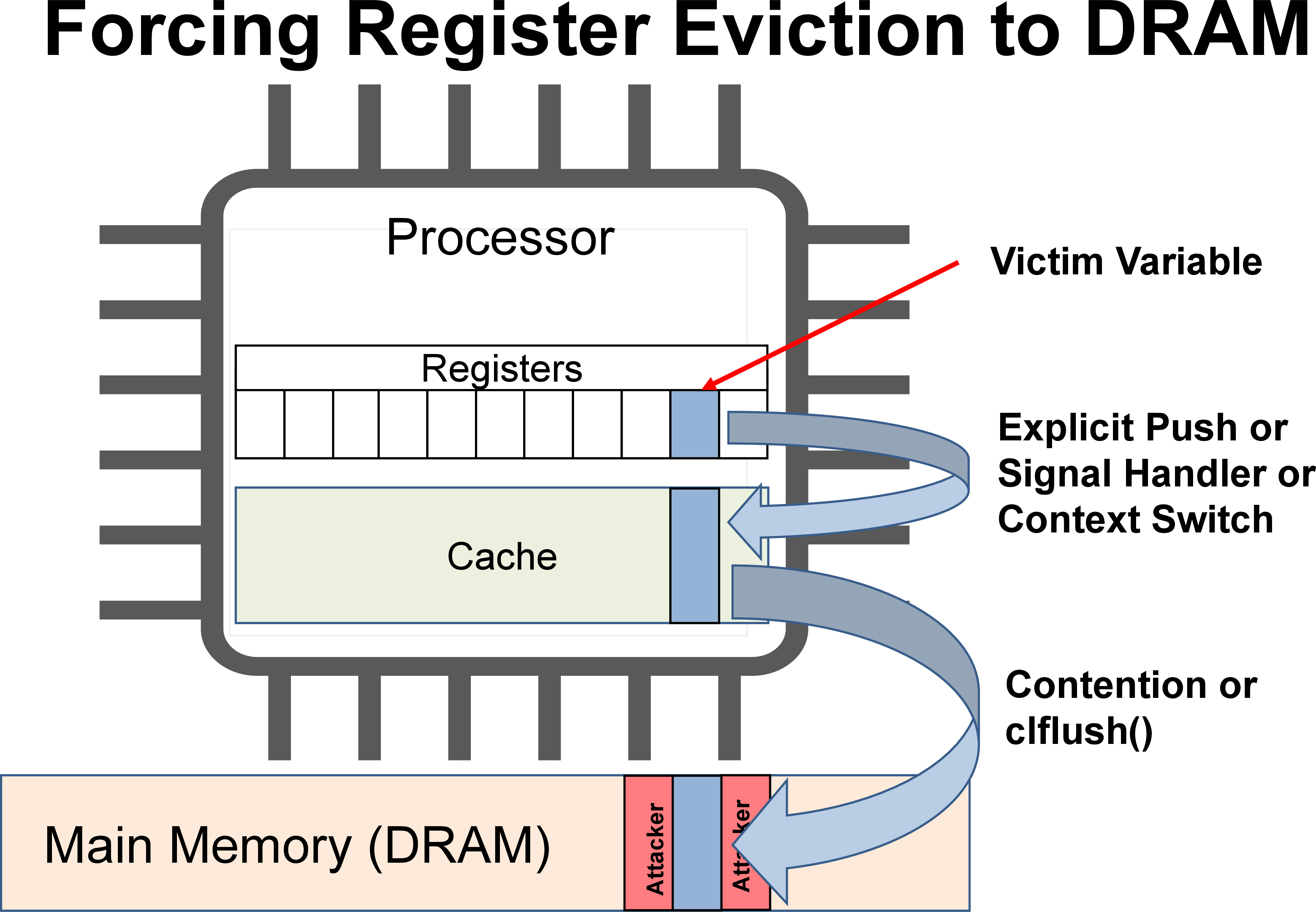}
    \caption{We can evict registers to stack by switching contexts, which pushes the registers to cache, and then with contention, we can evict them to DRAM where data can be flipped with Rowhammer.}
    \label{fig:register_eviction}
\end{figure}

There are a number of common functions that push register values to stack by default, as shown in Figure \ref{fig:register_eviction}. For example, the \texttt{ebx} register is pushed to stack by both the \texttt{sleep()} function \cite{sleep3} and the \texttt{getchar()} function \cite{getchar3p}. Listing \ref{lst:example_push} from the \texttt{glibc} library shows the \texttt{recv} function pushes the \texttt{ebx} register to stack, and pops the \texttt{ebx} register after the function completes. This is a common convention because registers are a fast but limited resource, and values are pushed and popped from stack to optimize their usage. Once such a code pattern of storing security variables in registers and then pushing them to stack is found, these values can be attacked via Rowhammer.

\label{section:register_attack}
\paragraph{Active}  We can actively force registers into the kernel stack by triggering a signal handler function that pushes the registers into the kernel stack. This is a built-in part of the Linux kernel to optimize register usage. This enables a new type of active attack, where we can target variables that are stored in registers. As seen in Figure \ref{fig:register_eviction}, even though the variables may not be stored in the stack during the compilation convention (as discussed previously), we can send a signal to the victim process or create a contention by running another process or making a system call. This will result in the victim process storing its CPU registers in the kernel stack, making the variables vulnerable to a Rowhammer attack.
We found experimentally that signal handlers implemented in the C programs by default push registers 
to stack, so any vulnerable data stored in those registers would be candidates for a Rowhammer attack. 

Since SIGSTOP cannot be handled by user programs, using it alone will not flush registers to the user stack. However, if a program has a custom signal handler\footnote{PoC register spilling using SIGINT: https://gist.github.com/anonymous-60819/d1d6137e17c34f761f7b33d60e922c9d}, the Linux kernel saves the register content to the user stack while the signal handler is being executed. This mechanism does not rely on the user's \textit{application logic}. When we send SIGSTOP right after the previous signal, the user’s signal handler also stops, and the register content of the user program stays in memory until receiving a SIGCONT, giving the attacker time to execute Rowhammer.

This is different than context switching which may force registers into kernel stack. 
In the context-switching case, the OS schedules each process for a specific amount of time, switching between them as needed. The OS saves the contents and state of the CPU (including registers) to a stack in order to allow a process to resume from where it left off when it is reloaded. The contents of some registers are saved to the kernel stack associated with the process, which can still be flipped as shown in ~\cite{tobah2022spechammer}. 

%% file: sections/experiment_results.tex
\section{Experimental Evaluation}\label{sec:experiments}
\subsection{Experiment Setup}
The experiments are conducted on a system with Ubuntu 20.04.01 LTS with 5.15.0-58-generic Linux kernel installed. The system uses an Intel Core i9-9900K CPU with a Coffee Lake microarchitecture. We used a dynamic clock frequency rather than a static clock frequency to improve the practicality of the attack.
End-to-end attack experiments are done on a single DIMM Corsair DDR4 DRAM chip with part number CMU64GX4M4C3200C16 and 16GB capacity. DRAM row refresh period is kept as 64ms which is the default value in most systems. For the experiments on \texttt{sudo}, we use version 1.9.12p1~\footnote{\texttt{sudo} git commit number 3396267291328eccfcbc7bfb1729c77f30216513}, which is the latest \texttt{sudo} version at the time of this work. We use the portable \texttt{OpenSSH} library with version 9.1p1~\footnote{\texttt{OpenSSH} git commit number 0ffb46f2ee2ffcc4daf45ee679e484da8fcf338c} for SSH experiments. To better accommodate the server environment and reduce the noise caused by desktop applications, we use the OS in console mode. For Rowhammer to successfully attack the stack of a program, the variables being attacked need to be loaded into memory at the right time. For experimental purposes in \texttt{SUDO}, \texttt{OpenSSH}, \texttt{OpenSSL} and \texttt{MySQL}, we used signals to make sure that the programs were synchronized. 

\subsection{Reproducibility of Bit Flips}
\label{sec:reproducibility}

\begin{figure}
  \subfloat[DDR3]{
	\begin{minipage}[c][1\width]{
	   0.25\textwidth}
	   \centering
	   \includegraphics[width=1\textwidth, trim={0 0 0 16mm},clip]{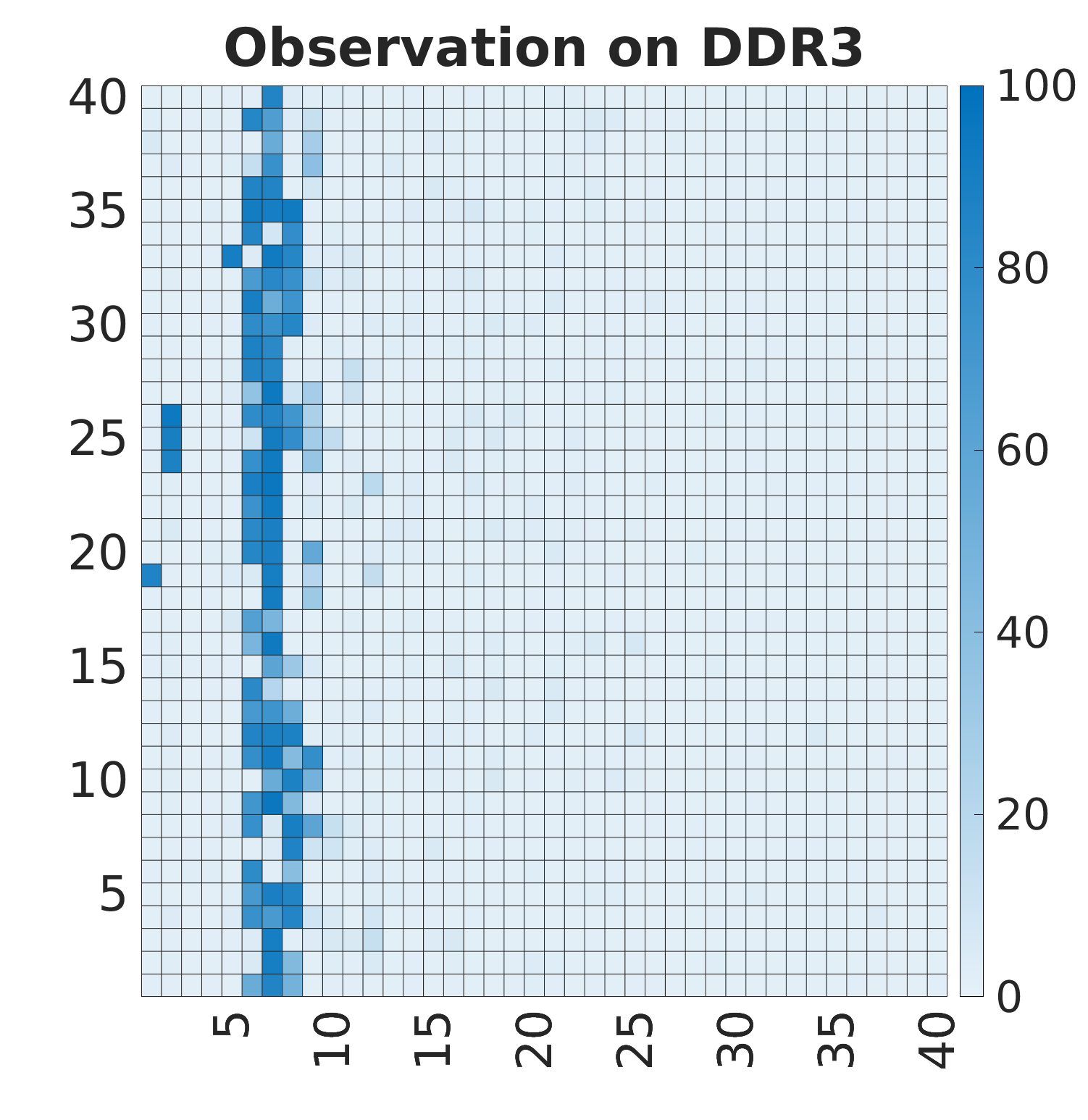}
	\end{minipage}}
  \subfloat[DDR4]{
	\begin{minipage}[c][1\width]{
	   0.25\textwidth}
	   \centering
	   \includegraphics[width=1\textwidth, trim={0 0 0 16mm},clip]{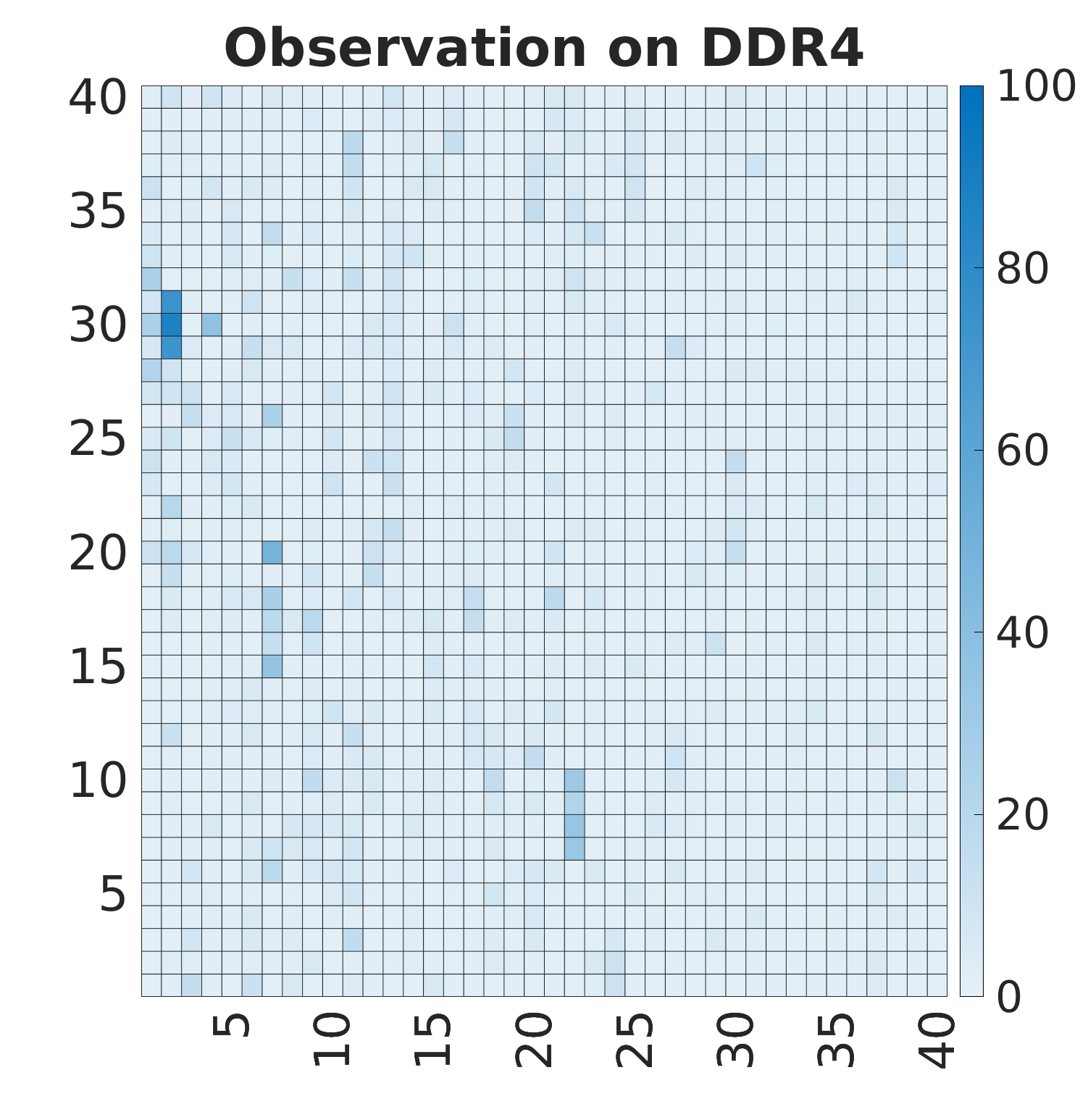}
	\end{minipage}}
        \caption{The comparison of heat maps of bit flips in DDR3 and DDR4 DRAM chips. Darker color illustrates the locations of more reproducible bit flips. The bit flips seen in DDR4 are less reproducible than DDR3.}
\label{fig:heatmap}
\end{figure}

Until this work, the reproducibility of bit flips induced by Rowhammer was not analyzed in detail. Therefore, it was not known whether each flippy location has equally reproducible or not. As the target size gets closer to a page size, every bit flip found is potentially useful since it will land on the target. 
However, as the target requires more precision, it is harder to find aligned bit flips; therefore, it is critical to attempt only when we find highly reproducible bit flips. This way, we can put the burden on the offline memory profiling phase and keep the online time as short and accurate as possible.
To test the reproducibility of bit flips, we select a 64 MB physically contiguous memory buffer. In DDR3, we apply double-sided Rowhammer and slide the Attacker-Victim-Attacker window by one at every step. Once we finish the buffer, we store the bit flip locations and start the same process from the beginning. We hammer the same memory buffer for 100 times and count the number of flips for each bit location that has flipped at least once. We found 1667 unique flippy bit locations in total. Figure~\ref{fig:heatmap} illustrates the frequency of bit flips in a heat map. 
We observe that only a limited portion of found bit flips are actually reproducible, while most of them are not reproducible at all in 100 trials. 
\raggedbottom

\subsection{Success Rate of Baiting Method}\label{sec:success_rating}

During our experiments, we found that with ASLR enabled, we could successfully locate the target page into the flippy row about 30\% of the time. With further engineering efforts, this number can be brought up to 80\% \cite{kwong2020rambleed}. In Section~\ref{sec:attacks}, we refer to the ability to locate our target page into the flippy row as our bait-page success rate, as we deallocate a set number of bait pages for the system in the hopes it will force our target variable into a vulnerable place in memory. 
We also observed that ASLR-randomized page offsets can partially leaked through the number of Page Faults which can further optimize our attack as seen in Figure~\ref{fig:page_faults}. Further analysis is given in Appendix~\ref{sec:bypass_aslr}.

\begin{figure}
    \centering
    \includegraphics[width=\columnwidth]{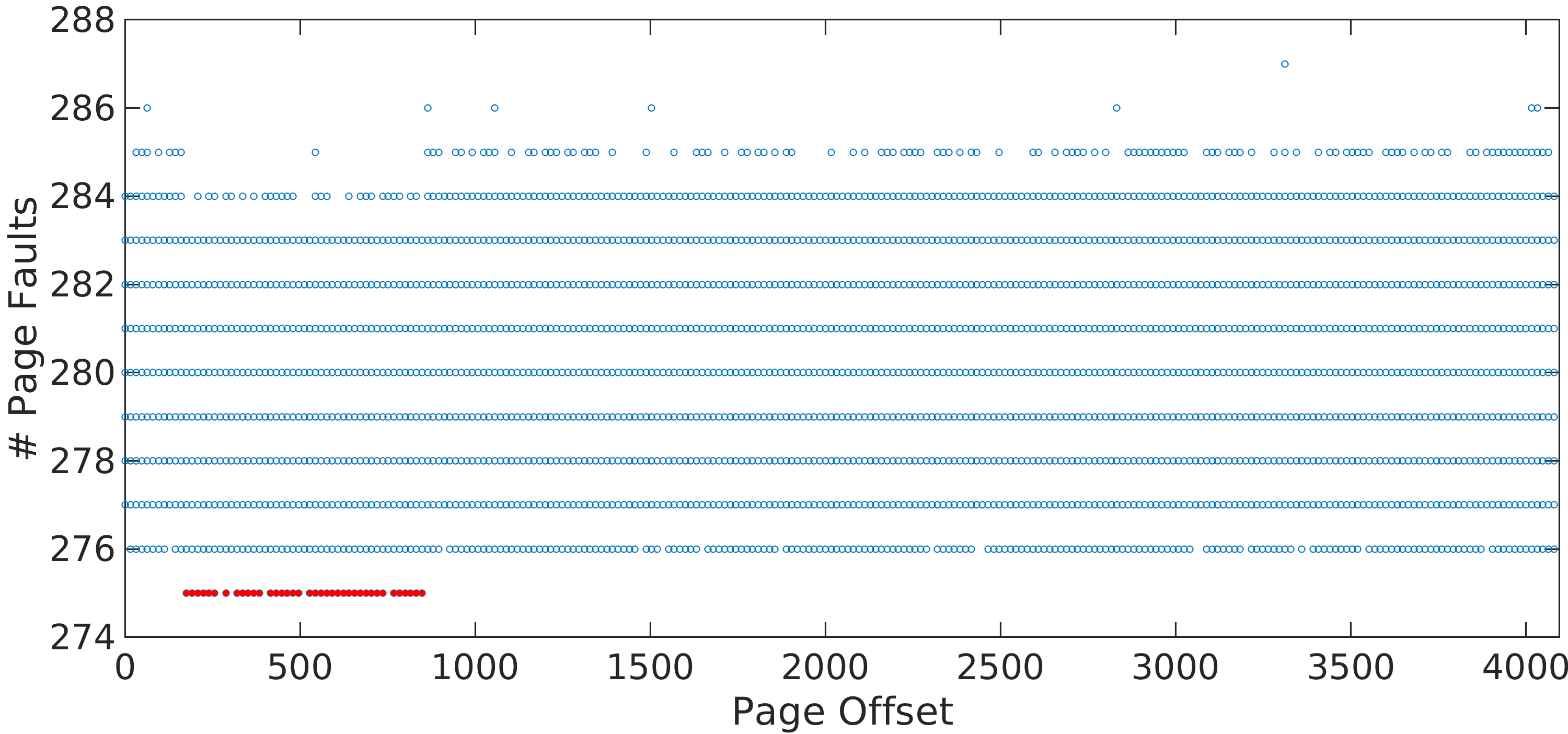}
    \caption{Page Fault Side Channel Analysis Demonstrating A Relationship Between Minor Page Faults and Page Offset}
    \label{fig:page_faults}
\end{figure}

\subsection{Evaluation on Different DRAM Chips}
Both offline and online phases of our attack require finding bit locations that are vulnerable to Rowhammer attack. Since the bit flip frequency depends heavily on how flippy a DRAM chip is, we evaluate our attack on different DRAM chips from both DDR3 and DDR4 memory profiles. We have taken 14 DDR3 memory profiles from~\cite{tatar2018hammertime}, and we generated the remaining 6 memory profiles on our DRAM chips. In total, we have analyzed 20 DRAM chips. The results are summarized in Table~\ref{tab:flip_profiles}.
In the last column, we can see the probability of finding at least one flip in a 32-bit integer after profiling 0.1\% of the total memory. To calculate this probability, we first find $n_{avg}$, the average number of bit flips that land on a 32-bit variable for 256 possible page offsets. Then, we calculate the probability of having a successful attack with a single flippy page by dividing the average flip count, $n_{avg}$ by the total number of flippy pages, $n_{\mathrm{flippy}}$. Finally, for a stealthy attack, we assume we only use 0.1\% of the total memory size, $N_{\mathrm{pages}}$.
The final fault probability is calculated as $p_{\mathrm{fault}}= (1 - (1 - n_{avg}/n_{\mathrm{flippy} })^{N_{\mathrm{pages}}/1000})\times100$.
While probabilities are over 90\% for most DRAM chips, it is important to note that other factors affect the probability of seeing a flip in the target variable of an actual process, including the probability that the process gets loaded into the flippy page in the first place. 
\begin{table}
\footnotesize
        \centering
    \begin{tabular}{c|cccc}
    \toprule
             & Brand & Serial Number & \thead{Size\\{[GB]}} & \thead{$p_{fault}$\\ {[\%]}} \hspace{-0.08in}\\
    \midrule
             \multirow{14}{*}{\rotatebox[origin=c]{90}{DDR3}} 
             &  Corsair & CMD16GX3M2A1600C9 & 16 & 99.99 \\ 
             & Corsair & CML16GX3M2C1600C9 & 16 & 99.99 \\  
             & Corsair & CML8GX3M2A1600C9W & 8 & 99.99 \\  
             & Corsair & CMY8GX3M2C1600C9R & 8 & 97.26 \\ 
             & Crucial &  \hspace{-0.08in}BLS2C4G3D1609ES2LX0CEU \hspace{-0.1in} & 8 & 72.34  \\  
             & Geil & GPB38GB1866C9DC &  8  & 99.95 \\ 
             & Goodram  & GR1333D364L9/8GDC & 8  & 57.47 \\ 
             & GSkill & F3-14900CL8D-8GBXM & 8  & 90.44   \\ 
             & GSkill & F3-19200C10-8GBZHD & 8 & 99.99 \\ 
             & GSkill & F3-14900CL9D-8GBSR & 8   &  88.76  \\ 
             & Hynix &  HMT351U6CFR8C-H9 & 8  & 99.77 \\ 
             & V7 & V73T8GNAJKI & 8  & 45.17 \\ 
             & PNY &  MD8GK2D31600NHS-Z & 6  & 92.58  \\ 
             & Integral & IN3T4GNZBIX & 4  & 79.19 \\ 
             & Samsung & M378B5173QH0 & 4  &  69.67 \\ 
             & Samsung & M378B5773DH0 & 2  &  99.69 \\ 
    \midrule
            \multirow{4}{*}{\rotatebox[origin=c]{90}{DDR4}} 
             & Corsair &CMU64GX4M4C3200C16&64& 99.99  \\ 
             & Corsair &CMK32GX4M2B3200C16&32&  99.98  \\ 
             & GSkill &F4-3600C16D-16GVKC&16&  99.99   \\ 
             & Crucial &CT8G4DFD824A.C16FF& 8 &  90.47   \\ 
        \bottomrule
        \end{tabular}
    \caption{The probability of flipping at least one bit in a 32-bit integer calculated on 16 different DDR3 chips and 4 DDR4 chips per profile (128 or 256MBs). In our setup, it takes 95 minutes to profile a 128 MB on DDR3 and 480 minutes to profile 256MB on DDR4 chips.}
    \label{tab:flip_profiles}
\end{table}

\section{Attacks -- Injecting Faults into Programs}\label{sec:attacks}

Our attacks require finding vulnerabilities in the code we call \emph{Rowhammer gadgets}. Rowhammer gadgets are pieces of code with security-critical logic that can be corrupted and bypassed by a Rowhammer attack. It generally consists of a stack variable being set to an initial value, then changed depending on the program flow, and being evaluated as being \emph{not} equal to a certain value as illustrated in Listing \ref{lst:Rowhammer_gadget}.
We can define an integer stack variable \texttt{auth} as equal to zero initially, then after a password check (which would set \texttt{auth} to 1 if entered correctly), check if the variable is \emph{not} equal to zero. 
We would consider this example a Rowhammer gadget because any bit flip in the \texttt{auth} variable would result in it being not equal to 0, thus passing the authentication. It would be better for security-related code to require that code be equal to a certain value rather than check if it is not equal to a certain value. 
\begin{figure}
\begin{lstlisting}[frame=single,
                    language=C++,
                    label={lst:Rowhammer_gadget},
                    caption=Returns \texttt{AUTH\_SUCCESS} if password is correct \texttt{AUTH\_FAILURE} otherwise.]
// Gadget
int auth = 0;
//password check code
if(auth != 0)
	return AUTH_SUCCESS;
else
	return AUTH_FAILURE;
\end{lstlisting}
\end{figure}

\subsection{Bypassing SUDO Authentication}

\texttt{sudo} is a process in Linux-based operating systems that stands for Super User Do. It allows a user to obtain root access to reading, writing, and executing protected files given they enter the correct password. Breaking the functionality of \texttt{sudo} is a textbook privilege escalation attack and can be devastating to systems that hide crucial infrastructure behind the root password. 
The system administrator sets a root password that is stored and hashed on the system, and when a user enters a password, the hashes of the two passwords are compared, and if they match, root access is granted to the user. This is seen in the code sample given in Listing \ref{lst:matched_var}.

A fault injection attack has been proposed on the \texttt{sudo} program before using a different technique \cite{gruss2018another} that requires a specific bit flip. The researchers found areas in the \texttt{sudo} binary where a bit flip could result in an opcode change which could result in privilege escalation. The researchers found a total of 29 bits that could be flipped, resulting in privilege escalation. 
An opcode flip requires high precision; once a page with flippy bits is found through the Rowhammer profiling stage, a flippy bit needs to be located in the correct position within the page. Flip a bit that is a single bit-distance away from the target will result in a broken \texttt{sudo} program and may require up to a system reboot. In contrast, our attack on the Rowhammer gadget code works if any bit in the \texttt{matched} variable is flipped, consisting of 4 bytes or 32 bits for the \texttt{matched} variable alone. 


\begin{figure}
\begin{lstlisting}[frame=single,
                    label={lst:matched_var},
                    caption= Password authentication function in \texttt{sudo}. Returns \texttt{AUTH\_SUCCESS} if password is correct \texttt{AUTH\_FAILURE} otherwise.]
int sudo_passwd_verify(...) {
  char des_pass[9], *epass;
  char *pw_epasswd = auth->data;
  size_t pw_len;
  int matched = 0;
...
  epass = (char *) crypt(pass, pw_epasswd);
  if (epass != NULL) {
    if (HAS_AGEINFO(pw_epasswd, pw_len) 
	   && strlen(epass) == DESLEN)
      matched = !strncmp(pw_epasswd, epass, DESLEN);
    else
      matched = !strcmp(pw_epasswd, epass);
  }

  explicit_bzero(des_pass, sizeof(des_pass));

  debug_return_int(matched ? AUTH_SUCCESS 
                                    : AUTH_FAILURE);
}
\end{lstlisting}
\end{figure}



After running the \texttt{sudo} experiment for 10 hrs 34 minutes, we saw a total of 11 successful attacks where we gained root access.
This amounts to an average of about an hour of profiling, as seen in Table \ref{tab:all_results}, to see a successful attack. Additionally, we see that the total online time is less than an hour to see the 11 flips, so a total of 5 minutes of hammering on average on the \texttt{sudo} program itself to see a flip. 
The time between successful attacks occasionally varied - sometimes we would see 2-3 attacks in a 15-20 minute window of profiling. Other times it may take up to a few hours. We speculate this to be due to where the process is being placed in memory, as some areas of the DRAM banks may be more flippy than others due to manufacturing defects. 
We also noted that of the 5334 attacks, we saw 1989 attacks where the target variable was placed correctly in the flippy page. This is a bait-page success rate of about 37\%.

We were initially concerned that the Rowhammer would flip too many bits in the stack of the process that it would be unable to finish execution. While we did find that it was flipping bits in other variables other than \texttt{matched} unintentionally, the program still executed successfully, and when \texttt{matched} was flipped we gained root access. Fortunately, stability did not become an issue in our experiments. 
The results of the experiment demonstrate the novel attack on stack can enable privilege escalation by flipping bits in the stack.

\fussy
\subsection{Bypassing OpenSSH Authentication}

To demonstrate the extent of the new attack surface that our attack work enables, we implement the attack on SSH protocols. 
SSH (Secure Shell Protocol) is an application layer protocol that allows secure remote user login, command execution, and other remote network operations such as TCP port forwarding, tunneling, and file transfer. SSH protocol works in a client-server model. Public-key encryption is used for authenticating the client and the server to each other. After the authentication phase, the transferred data is secured using symmetric key encryption schemes, such as AES. Several libraries implement SSH protocol. OpenSSH is one of the most popular implementations of SSH protocol.
Several attacks on OpenSSH have been shown to steal RSA session keys~\cite{kwong2020rambleed}.

When the server program starts, it constantly monitors the incoming connections request to port 22 by default. This monitoring is achieved in an infinite \texttt{while} loop. When the server gets a connection request from a client with a given username and password, a chain of functions is called to check if the provided password is correct. Here, we mention the two most important ones that we can use for our attack. The first function is \texttt{mm\_answer\_authpassword}, and the second function is \texttt{auth\_password} which is called by the first one. We show the truncated versions of these functions in Listing~\ref{lst:openssh_func1} and \ref{lst:openssh_func2}.
Within these two functions, there are two different local variables that carry the information regarding if the user will be authenticated later on.

\begin{figure}
\begin{lstlisting}[frame=single,
                    label={lst:openssh_func1},
                    caption = Password authentication function in OpenSSH. Returns 1 if the password is correct and 0 otherwise.
]
int mm_answer_authpassword(...){
  char *passwd;
  int r, authenticated;
...
  authenticated=options.password_authentication 
        && auth_password(ssh, passwd);
...
  if ((r=sshbuf_put_u32(m, authenticated)) != 0)
	fatal_fr(r, "assemble");
...
  return (authenticated);
}
\end{lstlisting}
\end{figure}

\begin{figure}
\begin{lstlisting}[frame=single,
                    label={lst:openssh_func2},
                    caption= Password authentication function in OpenSSH. Tries to authenticate the user using password.  Returns true if authentication succeeds.]
int auth_password(...){
  Authctxt *authctxt = ssh->authctxt;
  int result, ok = authctxt->valid;
...
  if (*password == '\0' && options.permit_empty_passwd == 0)
    return 0;
...
  result = sys_auth_passwd(ssh, password);
  if (authctxt->force_pwchange)
    auth_restrict_session(ssh);
  return (result && ok);
}
\end{lstlisting}
\end{figure}

In function~\texttt{mm\_answer\_authpassword}, \texttt{authenticated} flag is set if the \texttt{auth\_password} returns \texttt{1} in line 5 and then returned. After being returned, the return value is checked if it equals \texttt{1}. The client is authenticated, and if the condition is met and the SSH session starts. Otherwise, the client is asked to enter the password again. If the correct password is not given in three trials, the client has to send the connection request again.
Here, the \texttt{authenticated} flag is stored in the stack memory of the program and, therefore, in DRAM, and a potential target for our attack. If we flip the least significant bit of this 32-bit integer value after line 5, we see that the client is authenticated regardless of the password value, and remote shell access is given. However, flipping other bits other than the least significant bit results in authentication failure, even if the password is correct.

The other target for our attack is the \texttt{result} flag in \texttt{auth\_password} function. It is initialized to \texttt{0} in line 3 and set to \texttt{1} in line 5 if the password is correct. Note that the \texttt{result} flag is given to a \textit{logical and} operation with \texttt{ok} flag. \texttt{ok} flag is set to \texttt{1} if username is valid. Therefore, as long as the \texttt{result} is a nonzero value, the return value would be \texttt{1}. This logic increases the changes of our attack since as long as we flip any bit of the 32 bits of the \texttt{result} variable, we can successfully bypass the password authentication.

Table~\ref{tab:all_results} shows the averages of a successful attack on the \texttt{result} variable in SSH. We observed that over the course of about one and a half hours, we saw two total successful logins into the SSH server without the correct password, which would be an average of 45 minutes, as seen in Table \ref{tab:all_results}. This required a total of 11 minutes of online time, for an average of about 6 minutes of hammering SSH per successful attack. In order to complete the attack, we found 1025 memory pages in the system with flippy bits. 
We also saw that of the attacks, 412 out of 1025 released the correct number of bait pages such that the target variable of SSH was placed correctly in the flippy row. This is a bait-page success rate of about 40\%.

\subsection{Attack on \texttt{OpenSSL} Security Checks Stored in Stack}
We experiment with a simple \texttt{OpenSSL} process where we target a security check variable. At the end of the ECDSA sign setup method, a security check determines if a variable called \texttt{ret} is not equal to zero. If the variable is equal to zero, it means that a security check failed and a jump occurred past where the variable is set to \texttt{1}, indicating all security checks passed. A successful security bypass would hammer the security variable in the stack and force it to be \texttt{1} regardless of if it made a jump or not. This could potentially be used in conjunction with a Rowhammer attack that targets dynamic memory. 


From Table \ref{tab:all_results} we can see that there is an average offline time of 1 hr 45 mins, and an average online of 7 minutes. This required only 14 minutes of hammering on OpenSSL itself. During profiling, 1372 pages were found to be flippy, and 277 of them were correctly utilized by having the target security variable placed in them during the attack stage, a resulting bait-page success rate of about 20\%.

\begin{table}
\centering
    \begin{tabular}{l|c c c}
    \toprule
        Category  & SUDO & OpenSSH & OpenSSL \\
    \midrule
        Total Time  & 1 hr 9 mins & 45 mins & 1 hr 45 mins \\ 
        Online Time  & 5 mins & 6 mins & 7 mins\\ 
        Flippy Pages  & 485 & 513 & 686\\ 
        Correct Baiting  & 181 & 206 & 139 \\ 
    \bottomrule
    \end{tabular}
    \caption{Results from the \texttt{SUDO}, \texttt{OpenSSH} and \texttt{OpenSSL} experiments showing offline time and online time, and the number of flippy pages found, as well as the number of attacks with the correct number of bait pages released}
    \label{tab:all_results}
\end{table}

\section{Vulnerability Analysis}\label{sec:vuln_analysis}

\subsection{RSA Bellcore Attacks}
The early work by Boneh et al.~\cite{Boneh2015OnTI} popularly referred to as Bellcore attacks, demonstrated the importance of checking for errors in cryptographic implementations in a CRT-based RSA implementation. The first mitigation against Bellcore attacls on OpenSSL was released in 2001 and is shown in Listing~\ref{lst:bellcoreopenssl}
\footnote{\url{https://github.com/openssl/openssl/commit/1777e3fd5eac0e491bb16a0bb37f4b0f298e6486}} 
The current OpenSSL implementation performs a check operation to find if an error occurred after the fast CRT-based RSA exponentiation. If an error is detected, the code runs a slower (non-CRT based) exponentiation to compute the signature, thus preventing the possibility of initiating the Bellcore attack. The check mechanism involves recomputing the message using the signature and public key. Recomputed message is later subtracted from the original message to check if both are the same message. If the result is zero, it means the messages match and the exponentiation is computed correctly. The zero check function can be seen in line 17 in Listing~\ref{lst:bellcoreopenssl}.    

For a successful attack, the first step is to create a fault in one of the partial CRT-based RSA computations. Then, another fault is introduced in the check mechanism to trick the code into thinking the CRT-based RSA exponentiation has been calculated correctly\footnote{The probability of both faults going through will be low, however Bellcore requires only one faulty sample to succeed.}. This is achieved by launching a stack attack on the function $BN\_is\_zero$. When line 17 calls for $BN\_is\_zero$ function, the result of the zero check is returned using the \texttt{EAX} register. We can force the process to halt and put the result on the stack. By using Rowhammer, we can manipulate the variable once it is on the stack. When the return value is anything other than zero, the if case will be executed, giving the appearance that the CRT-based exponentiation was computed correctly.   

\begin{figure}
\begin{lstlisting}[frame=single,
                     label={lst:bellcoreopenssl},
                    caption= Error checking in OpenSSL ModExp]
static int rsa_ossl_mod_exp(BIGNUM *r0, const BIGNUM *I, RSA *rsa, BN_CTX *ctx){
...
  if (rsa->e && rsa->n) {
        if (rsa->meth->bn_mod_exp == BN_mod_exp_mont) {
            if (!BN_mod_exp_mont(vrfy, r0, rsa->e, rsa->n, ctx,
                                 rsa->_method_mod_n))
                goto err;
        } else {
            bn_correct_top(r0);
            if (!rsa->meth->bn_mod_exp(vrfy, r0, rsa->e, rsa->n, ctx,
                                       rsa->_method_mod_n))
                goto err;
        }...
        if (!BN_sub(vrfy, vrfy, I))
            goto err;
        if (BN_is_zero(vrfy)) {
            bn_correct_top(r0);
            ret = 1;
            goto err;   /* not actually error */
        } ...
}
 
\end{lstlisting}
\end{figure}

\subsection{Bypassing \texttt{MySQL} Authentication}
\texttt{MySQL} is the most popular open-source database management system~\cite{solid2018db}, which is commonly used by many organizations and websites in all industries from Defense \& Government to Social Networks including US Navy, NASA, Twitter, Facebook, LinkedIn, and Bank of America\cite{mysql2023customers}. 

We found a Rowhammer gadget~\footnote{\url{https://github.com/mysql/mysql-server}} given in Listing~\ref{lst:mysql_func} in the source code of \texttt{MySQL} server that is used for authenticating a client with a password. The password check happens between line 3 and 7 and the result is stored in \texttt{fast\_auth\_result}. 
When we simulate a 0 to 1 flip on \texttt{fast\_auth\_result.first} in line 8, we observe that the client is authenticated even with an incorrect password. Note that, unlike the previous attacks, the target variable requires single-bit precision; hence, the attack is harder to achieve using Rowhammer.

\begin{figure}
\begin{lstlisting}[frame=single,
                    label={lst:mysql_func},
                    caption= \texttt{MySQL} password authentication. Tries to authenticate the client using \texttt{authorization\_id} and \texttt{scramble}. The authentication succeeds if \texttt{fast\_auth\_result.first} is false.]
static int caching_sha2_password_authenticate(...){
...
  std::pair<bool, bool> fast_auth_result =
      g_caching_sha2_password->fast_authenticate(
          authorization_id, reinterpret_cast<unsigned char *>(scramble),
          SCRAMBLE_LENGTH, pkt,
          info->additional_auth_string_length ? true : false);
  if (fast_auth_result.first) {
    if (vio->write_packet(vio, (uchar *)&perform_full_authentication, 1))
      return CR_AUTH_HANDSHAKE;
  } else {
    if (vio->write_packet(vio, (uchar *)&fast_auth_success, 1))
      return CR_AUTH_HANDSHAKE;
    if (fast_auth_result.second) {
      const char *username =
          *info->authenticated_as ? info->authenticated_as : "";
    }
    return CR_OK;
  }
\end{lstlisting}
\end{figure}

%% file: sections/tls.tex
\section{End-to-End Attack Example}\label{sec:full_attack}
The effectiveness of this attack by demonstrated by deploying a client/server signature verification via OpenSSL. Note that this example does not use signals or signal handlers for synchronizing the attacker and the victim but rather uses the concept of a blocking window inherent to the client/server signature verification process to ensure the attacker hammers at the right time. The attacker is assumed to be colocated with the client and will hammer the victim client's high-level signature verification process forcing it to interpret a faulty signature as valid. This is in the context of a man-in-the-middle attack, where an attacker is trying to trick a client into thinking a server is the authentic target they are trying to connect to. 

In the typical scenario, the client will attempt to connect to the server and will send a \texttt{ClientHello} message to the server. The server will respond with a \texttt{ServerHello} message, which includes the server's public key and a signature of the handshake. The client will then verify the signature using the server's public key. If the signature is valid, the client will assume that it is safe to send sensitive information to the server. If the attacker can flip a bit in the signature verification process, the client will think the signature is valid and will send sensitive information to the attacker.

\begin{figure}[h]
    \centering
    \includegraphics[width=.85\columnwidth]{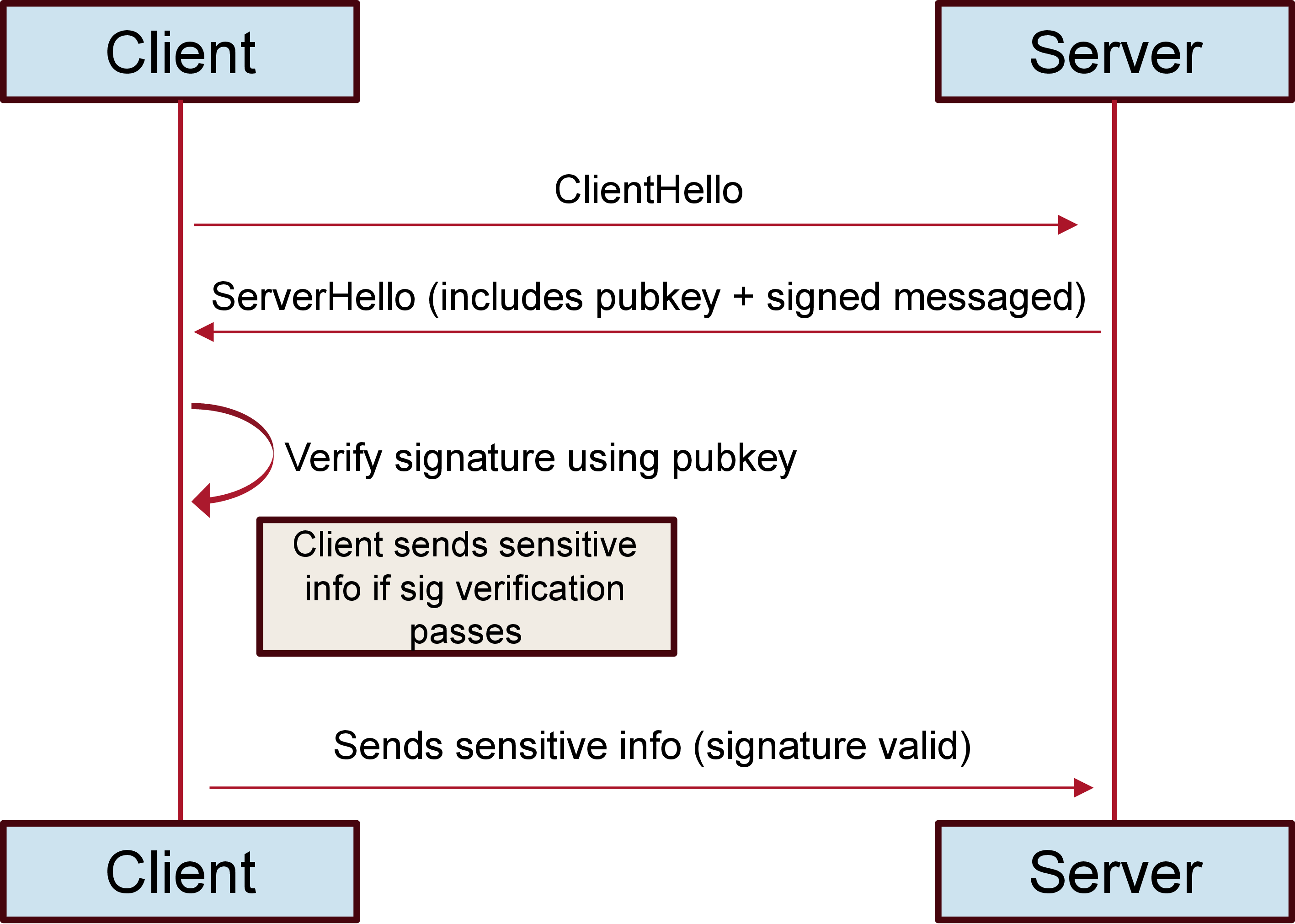}
    \caption{Typical scenario where the client connects to the server, sends a message and receives the message signed by the server and is able to authenticate the server.}
    \label{fig:poc_1}
\end{figure}

In Figure \ref{fig:poc_1}, we see the typical scenario where the client connects to the server, sends a message and receives the message signed by the server, and is able to authenticate the server. Importantly, the client is vulnerable to the Rowhammer attack during the phase while it is waiting for a response from the server. This connection phase can take time (in the order of milliseconds) and is ultimately controlled by the server, and therefore, the attacker can hammer the client's memory during this phase.

\begin{figure}
    \centering
    \includegraphics[width=.9\columnwidth]{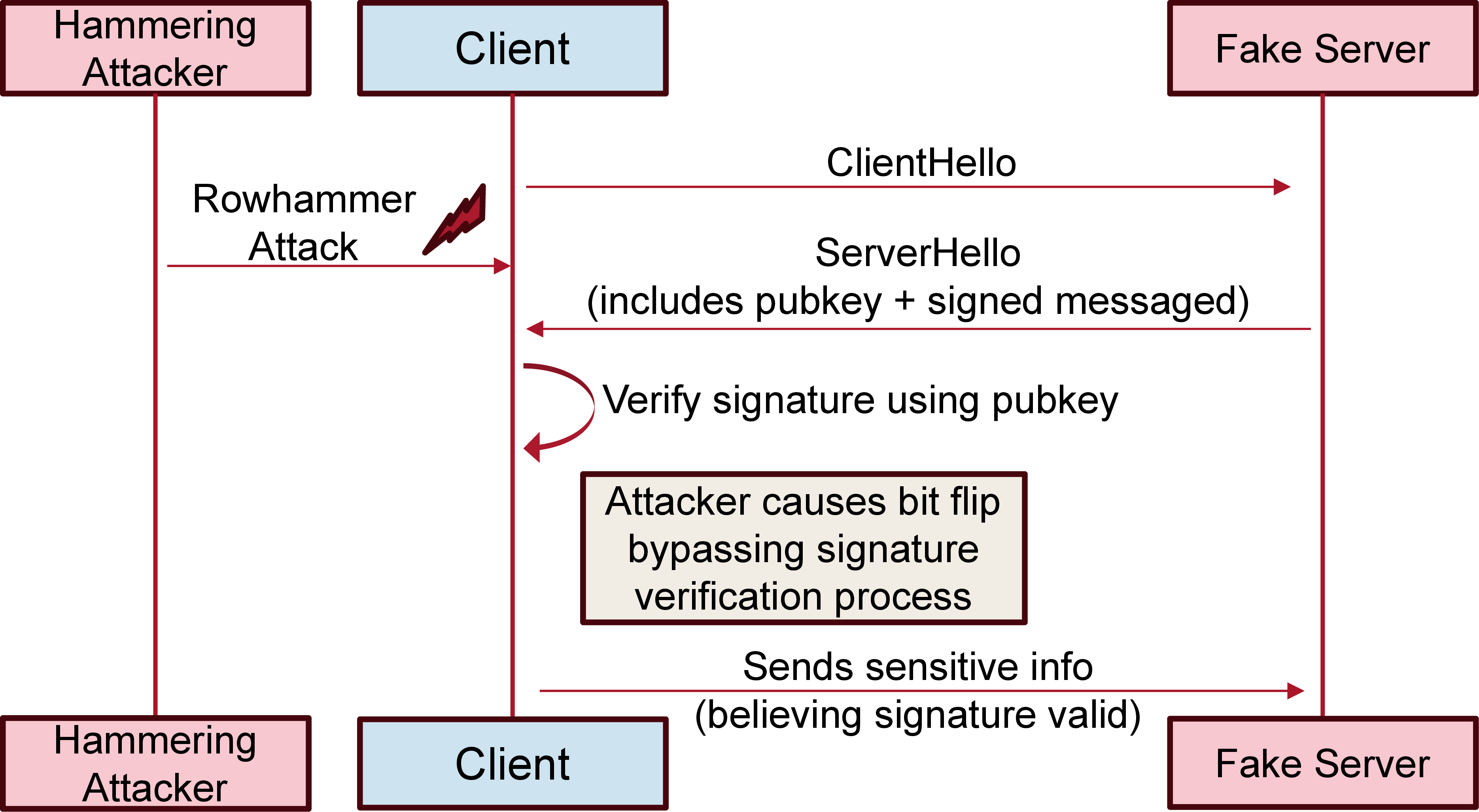}
    \caption{Attack scenario where the attacker acts as both the fake server and colocated with the client.}
    \label{fig:poc_2}
\end{figure}

In Figure \ref{fig:poc_2}, we can see the attack scenario. The attacker capitalizes on the fact that the client is vulnerable to Rowhammer during the connection phase. The attacker acts as both the server and is colocated with the client. The attacker responds to the clients \texttt{ClientHello} with a \texttt{ServerHello} message, which includes the attacker's public key and a signature of the handshake. The client will then verify the signature using the attacker's public key. If the attacker can flip a bit in the signature verification process, the client will think the signature is valid and will send sensitive information to the attacker. Theoretically, The attacker can then forward the message to the real server and receive the response. The attacker can then forward the response to the client, and the client will think it is communicating with the real server, otherwise known as a man-in-the-middle attack.

\begin{figure}
\footnotesize
    \begin{lstlisting}[frame=single,
                        language=C,
                        label={lst:client_code_1},
                        caption=Client code that connects to the server and sends a message to be signed. It is vulnerable to the Rowhammer attack during the connection phase.]
  int pass=0;
  // Create client socket
  client_fd = socket(AF_INET, SOCK_STREAM, 0);
...
  // Connect to server
  connect(client_fd, (struct sockaddr *)&server_addr, sizeof(server_addr));
  // Send a message to the server
  unsigned char message[32] = "message";
  send(client_fd, message, sizeof(message), 0);
...
  while ((bytes_received = recv(client_fd, buffer, sizeof(buffer), 0)) > 0){
      memcpy(sig_buf + sig_len, buffer, bytes_received);
      sig_len += bytes_received;
  }
  // Deserialize the signature
  const unsigned char *pp = sig_buf;
  ECDSA_SIG *signature = d2i_ECDSA_SIG(NULL, &pp, sig_len);
...
  // Verify the signature
    if (verify_message(message, sizeof(message), signature, ec_key)==SUCCESS){
        pass = 1;
    }
    \end{lstlisting}
\end{figure}

This full attack scenario consists of 3 steps:

\begin{itemize}[noitemsep,topsep=0pt,leftmargin=*]
    \item \textbf{Step 1:} The client connects to the attacker and sends a \texttt{ClientHello} message.
    \item \textbf{Step 2:} The attacker sends a \texttt{ServerHello} message to the client, which includes the attacker's public key and a signature of the handshake.
    \item \textbf{Step 3:} The client will then verify the signature using the attacker's public key. If the attacker can flip a bit in the signature verification process, the client will think the signature is valid and will send sensitive information to the attacker.
\end{itemize}

\subsection{Taking advantage of IP Sockets for Synchronization}\label{sec:IP_sync}

This attack does not require any degradation or other synchronization techniques to time the bit-flip attack on the client. This is because the attacker is controlling the time that the verification process takes, and thus can simply wait for the bit flip to occur before sending the response to the client. 

In Listing \ref{lst:client_code_1} we see that the client has the ability to verify a signature based on the public key. It keeps the state of the verification process in the variable \texttt{pass}. The \texttt{pass} variable is set to 1 if the signature is valid, and 0 otherwise. During the connection phase, the Rowhammer attacker can attack the \texttt{pass} variable and flip a bit to make the client think the signature is valid.

\begin{figure}[]
    \begin{lstlisting}[frame=single,
                        language=C,
                        label={lst:client_code_2},
                        caption=Client code that uses the \texttt{pass} variable to determine if the signature is valid.]
// remove sensitive data from memory
EC_KEY_free(ec_key);
ECDSA_SIG_free(signature);
...
if (pass != 0) {
    fprintf(stdout, "Server Authenticated\n");
}
    \end{lstlisting}
\end{figure}

In Listing \ref{lst:client_code_2} we can see that \texttt{pass} is used to verify if the server is authenticated. If \texttt{pass} is not 0, then the server is authenticated. The attacker can flip a bit in the \texttt{pass} variable to make the client think the server is authenticated regardless of the TLS signature verification process executed previously. 

Just as with the previous experiments, this full attack was conducted on a system with Ubuntu 20.04.01 LTS with 5.15.0-58-generic Linux kernel installed. The system uses an Intel Core i9-9900K CPU with a Coffee Lake microarchitecture. We used a dynamic clock frequency rather than a static clock frequency to improve the practicality of the attack.

\subsection{Flipping a Register Value Pushed to Stack}

The high level source code for the OpenSSL signature verification can be modified to seemingly make it more difficult to attack with Rowhammer. We can force \texttt{pass} to go to a register with the following C code from Listing \ref{lst:client_code_3}. 

\begin{figure}[!h]
\footnotesize
    \begin{lstlisting}[frame=single,
                        language=C,
                        numbers=none,
                        label={lst:client_code_3},
                        caption= The \texttt{pass} security variable is stored in a register instead of stack]
register int pass asm("rbx") = 0;
    \end{lstlisting}
\end{figure}
\raggedbottom

It is a common practice by compilers that register space is used by default when possible to increase performance, but the C code in Listing \ref{lst:client_code_3} makes it explicit. After assigning \texttt{pass} to register \texttt{rbx}, the code behaves the same, but during the blocking window when OpenSSL is waiting to receive data from the server, register \texttt{rbx} is pushed to stack where it can be attacked. This is also common practice to maximize the utilization of registers which are a limited resource. When the data is popped off the stack after receiving data from the server, if it has been corrupted by Rowhammmer and that corrupted data is then put into the register.

\begin{table}
    \centering
    \begin{tabular}{l|c c}
    \toprule
        Category & Stack & Register \\
    \midrule
        Total Time & 27 mins & 36 mins \\ 
        Online Time & 20 mins & 31 mins \\ 
        Total Flippy Pages & 447 & 402\\ 
        Total Attacks w/ Correct \# of Bait pages & 104 & 105 \\ 
    \bottomrule
    \end{tabular}
    \caption{Results from the end-to-end attack on code using OpenSSL client/server signature verification}
    \label{tab:all_results_openssl}
\end{table}

\subsection{Results from End-to-End Attack}
We were able to successfully force the client to misauthenticate the digital signature sent by the server. Table \ref{tab:all_results_openssl} summarizes the results. It is notable that the results for attacking the variable in the stack, and attacking it when it is pushed from a register are comparable from a practical standpoint. Also note that after finding a flippy location in memory, the stack variable or register variable was loaded into the correct address 23\% and 26\% of the time respectively. Based on these findings, we can conclude that Register variables are no longer safe against Rowhammer. 

%% file: sections/countermeasures.tex
\section{Countermeasures}\label{sec:countermeasures}
\subsection{System Changes to Prevent Rowhammer}
One of the most common countermeasures cited is increasing the DRAM refresh rate. 
A faster refresh rate will result in worse performance and more power consumption and is not an ideal solution. 
Although  various row refresh methods have been proposed to reduce the overhead such as parallel ~\cite {HiRA2022}, and probabilistic row refresh~\cite{Wang2021}, they are not yet available for use in consumer systems.

One possibility is Hidden Row Activation (HiRA). HiRA is a novel technique proposed in \cite{HiRA2022} which parallelizes row refreshes for DRAM. It allows a row refresh operation to be hidden in the background while a row in the DRAM is being accessed or refreshed in the same bank. It takes advantage of the fact that different rows in the same bank may be connected to different charge restoration circuitry, allowing for concurrent refreshes. By making an effort to reduce the latency of refresh operations, HiRA can reduce the time window for Rowhammer attacks. HiRA claims to be able to concurrently refresh 32\% of rows in a DRAM concurrently on 56\% of off-the-shelf DRAM chips. However, despite strides in the direction of more efficient refreshes as a rowhammer mitigation, HiRA is still in its infancy and is not yet available for use in consumer systems.

Additionally, \cite{Wang2021} proposes a novel and efficient Rowhammer mitigation by building on existing Probabilistic Adjacent Row Activation (PARA) Rowhammer defences by building Discreet-PARA. Discreet-PARA combines disturbance bin counting, a mechanism for managing refresh operations on rows likely to be corrupted by Rowhammer, and PARA-cache, which is a cache that stores the most recently accessed rows. By tracking accesses and refreshes to rows, Discreet-PARA can detect and mitigate Rowhammer attacks. Researchers were able to optimize these refresh and access tracking mechanism to reduce the performance overhead from averages from 10.5\%-6.6\% to 5.3\%-2.6\%. Still, this mitigation results in overhead that may not be ideal for consumer systems. 

It was initially thought that Error Correcting Code (ECC) would be an ample countermeasure to Rowhammer. However, ECC is not a sufficient countermeasure because it can be defeated with triple bit flips~\cite{cojocar2019ecc}. ECC is a common feature in servers, but it is generally not available in consumer DRAMs.
\raggedbottom

\subsection{Tighter, More Precise Logic}\label{sec:tight_logic}
We propose a set of countermeasures that can be used against a Rowhammer attack on the stack of a process. The easiest way to make an attack more difficult is to tighten the logic of the code and avoid using if-not-zero conditionals. 

In the first example seen in Listing \ref{lst:counter_measure_bad1}, if any single bit in the \texttt{matched} variable is flipped, the first statement becomes true. The Rowhammer attack is not always precise, so checking if \texttt{matched} is not equal to zero allows an attacker to flip any of the 32 bits that make up \texttt{matched}, and the passwords will seem to match. This is a very similar gadget to the one we found in the \texttt{sudo} program.

\begin{minipage}[b]{0.40\columnwidth}
\centering
\lstset{label=SliceExaple2,columns=flexible}
\begin{lstlisting}[frame=single,
                    language=C++, label={lst:counter_measure_bad1},
                    caption=Loose Logic Suspectable to Rowhammer Attack]
if(matched != 0)
   //passwords match
else
   //passwords don't match
\end{lstlisting}
\end{minipage}
\hspace{2mm}
\begin{minipage}[b]{0.40\columnwidth}
\centering
\lstset{label=SliceExaple,columns=flexible}
\begin{lstlisting}[frame=single,
                    language=C++, numbers=none, label={lst:counter_measure_good},
                    caption=Tight Logic Less Suspectable to Rowhammer Attack]
if(matched == 1)
    //passwords match
else
    //passwords don't match
\end{lstlisting}
\end{minipage}

In contrast, Listing~\ref{lst:counter_measure_good} is safer because Rowhammer is required to flip \textbf{only} the least significant bit; otherwise, the passwords still will not match. 
Requiring security-sensitive variables to be stored in registers over stack is not an effective countermeasure because, as seen in Section~\ref{section:register_attack}, registers can be flushed to memory using signal interrupts and can still be flipped. 
The code we found in \texttt{sudo} and SSH have vulnerable code that is susceptible to Rowhammer by changing logic for any flip in the 32-bit variable,  while MySQL requires a least significant bit flip. Additionally, it can be beneficial to use boolean variables over integers when possible to reduce the target size. 

\begin{lstlisting}[frame=single,
                        language=C++, ,xrightmargin=0pt, numbers=none, label={lst:counter_measure_strong},
                        caption=Specific pattern in the matched variable increases the number of bits that need to be flipped resulting in a fault-resistant logic.]
if(matched == 0x69d61fc8)
    //passwords match
else
    //passwords don't match
\end{lstlisting}

Additionally, the stack Rowhammer attack can further be prevented by requiring specific patterns for security sensitve checks so a single bit flip will not result in a security failure. For our example with the \texttt{matched} variable, we could require that the variable be set to a random set of bits that are not all zeros. This would require an attacker to flip all bits in the variable to that specific pattern for authentication which is far more difficult than flipping any single bit. It takes advantage of the fact that rowhammer is a blunt tool that is often inprecise.


Consider listing \ref*{lst:counter_measure_strong}. This code is far more secure than the previous examples because it requires a specific pattern in the \texttt{matched} variable. This pattern is a random set of bits that are not all zeros. In this case, the attacker would need to flip the \texttt{matched} variable to \texttt{0110100111...}, which includes 17 bit flips in precise locations along the variable. This is more difficult than flipping any single bit in the variable.

\subsection{Detecting Rowhammer Gadgets} 
We believe that Rowhammer gadgets may be an excellent domain for a machine learning algorithm to find and detect vulnerable pieces of code using natural language processing. Similar work has been done using machine learning to detect Spectre gadgets \cite{tol2021fastspec}. A dataset of Rowhammer gadgets could be derived from existing code by simulating Rowhammer flips in stack variables and checking if the process experiences a security failure. 
\raggedbottom

\subsection{Responsible Disclosure}
We informed the library authors regarding the vulnerabilities we identified. SUDO authors committed a series of patches\footnote{\url{https://github.com/sudo-project/sudo/commit/7873f8334c8d31031f8cfa83bd97ac6029309e4f}} to make the library more resistant against Rowhammer by using mitigation described in Section~\ref{sec:tight_logic}. We have issued CVE-2023-42465 for SUDO.

%% file: sections/appendix.tex
\section{Spoiler Timings}

\begin{figure}[h]
    \centering
    \includegraphics[width=\columnwidth]{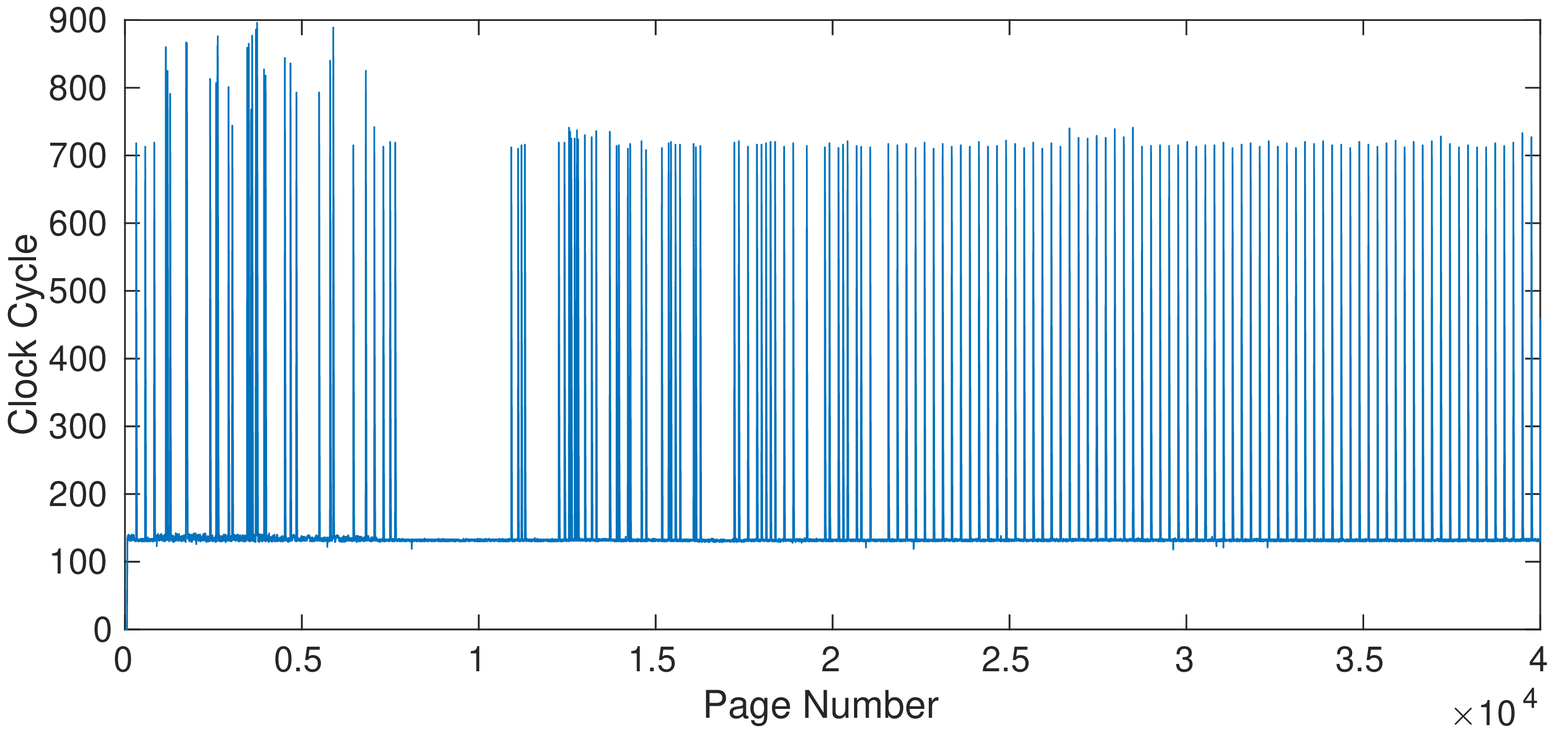}
    \caption{Timing peaks found by SPOILER. Equidistant peaks indicate physical continuity in memory.}
    \label{fig:spoiler_timing}
\end{figure}

The timings that are shown in \ref{fig:spoiler_timing} demonstrate how physical address dependencies result in more cycles because of speculative loading, as described in section \ref{offline_mem_profiling}

\section{Bypassing Stack ASLR}\label{sec:bypass_aslr}

\subsection{ASLR Background}
Address-space layout randomization (ASLR) is often used as a primary defense against memory corruption attacks. ASLR arranges the address space of a process randomly to prevent a user from targeting a specific area of code. It is supposed to rearrange the stack, heap, and libraries of an executable in a non-deterministic way. In theory, if an attacker finds a way to corrupt the memory it should not have access to, it should not be able to target any particular area in the process.

\begin{figure}
    \centering
    \includegraphics[width=\columnwidth]{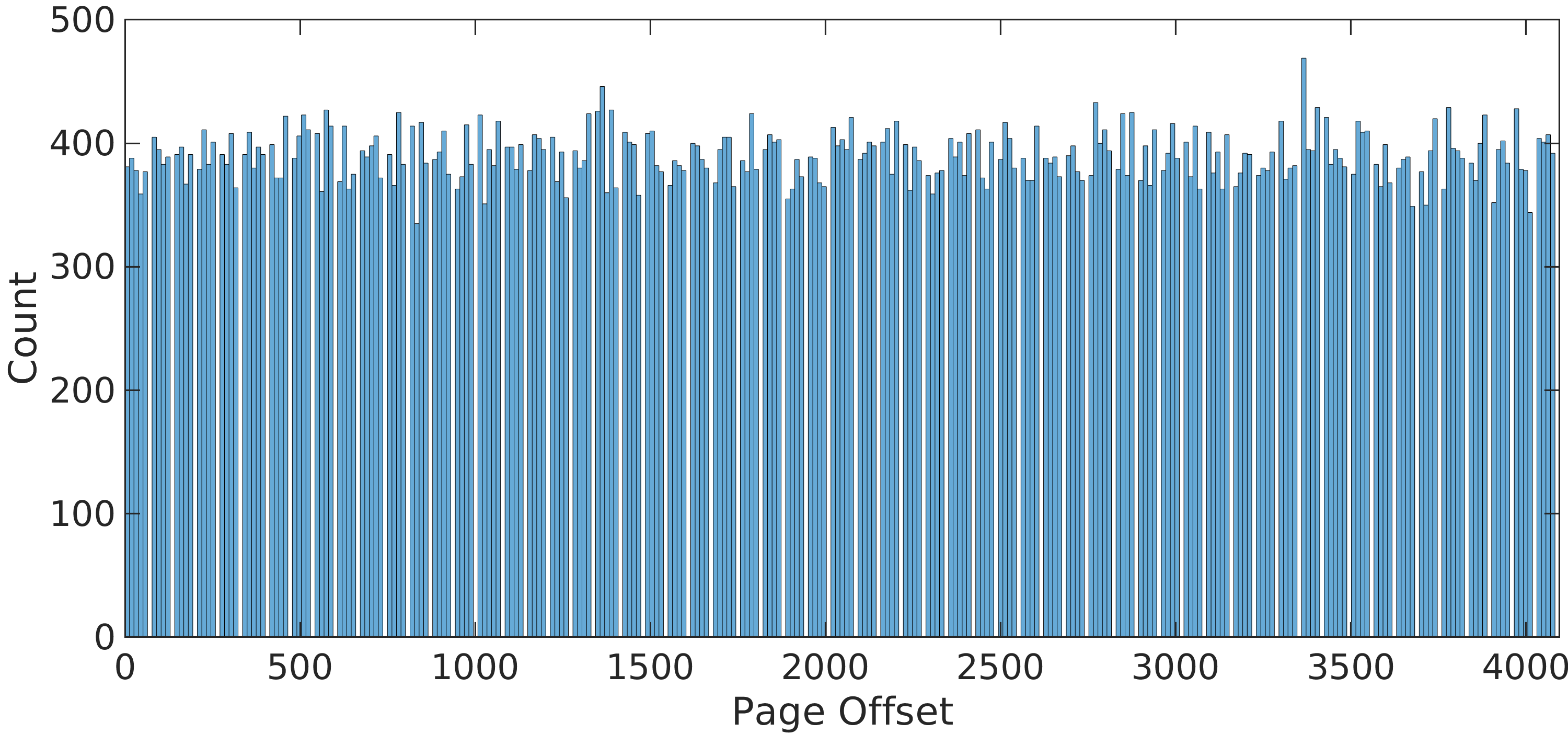}
    \caption{Histogram of page offset of a stack variable in stack memory out of 100K trials.}
    \label{fig:ASLR_random_vals}
\end{figure}

ASLR has been shown to be vulnerable in the past, typically through the use of software-side weak points such as memory disclosure vulnerabilities that reveal run-time addresses \cite{davi2015isomeron}. More recent attacks have also shown that ASLR can be broken through the use of EVICT+TIME cache attacks that can derandomize address spaces by correlating cache line addresses with page-table entries \cite{gras2017aslr}. Importantly, these attacks on ASLR do not circumvent stack ASLR, which is implemented in the Linux kernel as shown in Listing~\ref{lst:aslr_linux}. Stack ASLR should randomize the base address of the stack resulting in random variable offsets as seen in Figure \ref{fig:ASLR_random_vals} to the point a brute force attack randomly flipping bits in the system would be ineffective.
\begin{lstlisting}[ frame=single,
                    language=C++,
                    label={lst:aslr_linux},
                    caption= Page offset randomization for Stack memory in Linux Kernel]
unsigned long arch_align_stack(unsigned long sp){
   if (!(current->personality & ADDR_NO_RANDOMIZE) && randomize_va_space)
      sp -= get_random_int() % 8192;
   return sp & ~0xf;
}
\end{lstlisting}

Initially, it would seem that ASLR makes running a stack attack difficult. However, profiling a process to determine the number of bait pages required to be released to the system reduces the entropy significantly. 

The physical address is split into two parts; the page number and the page offset. The number of total bits in a physical address is calculated as ${\log_2 (p)}$ where $p$ is the total size of the main memory. For a system with 8 GB of main memory, the physical address is ${\log_2(8GB)}$ or 33 bits. Our operating system was also fragmenting memory into indivisible 4 KB-sized pages, which can be represented as 12 bits. This means that in the physical address of a system with 8 GB main memory and 4KB pages, the first 21 bits represent the page of the physical address, and the last 12 bits represent the offset within the page.

With our bait pages attack, we can effectively remove the entropy of the first 21 bits of randomization by forcing the base address to be placed on a known page around 45\% of the time, according to our findings. This leaves the last 12 bits of randomization to deal with. 
Through experimentation, we noticed that the entropy in the last 12 bits can be further reduced. We found that the last 4 bits of the address always stayed the same. When attacking \texttt{OpenSSL}, for example, we noticed that the last 4 bits always had a value of 0x8. If the variable we are attacking is a 4-byte variable, then there are only four possibilities for the last 4 bits of an address to be potentially vulnerable; $n, n+1, n+2$, and $n+3$, where $n$ is the starting address of the variable. This further reduces our ASLR entropy to a mere 8 bits, which can be easily exhausted.
We found a relationship between the number of bait pages required to be released by an attacker program to locate a variable in the Rowhammer page correctly and the \emph{offset} that variable appears within that page. We believe this to be a novel discovery because part of the intention of ASLR is to randomize the page offset.

\begin{figure}
    \centering
    \includegraphics[width=\columnwidth]{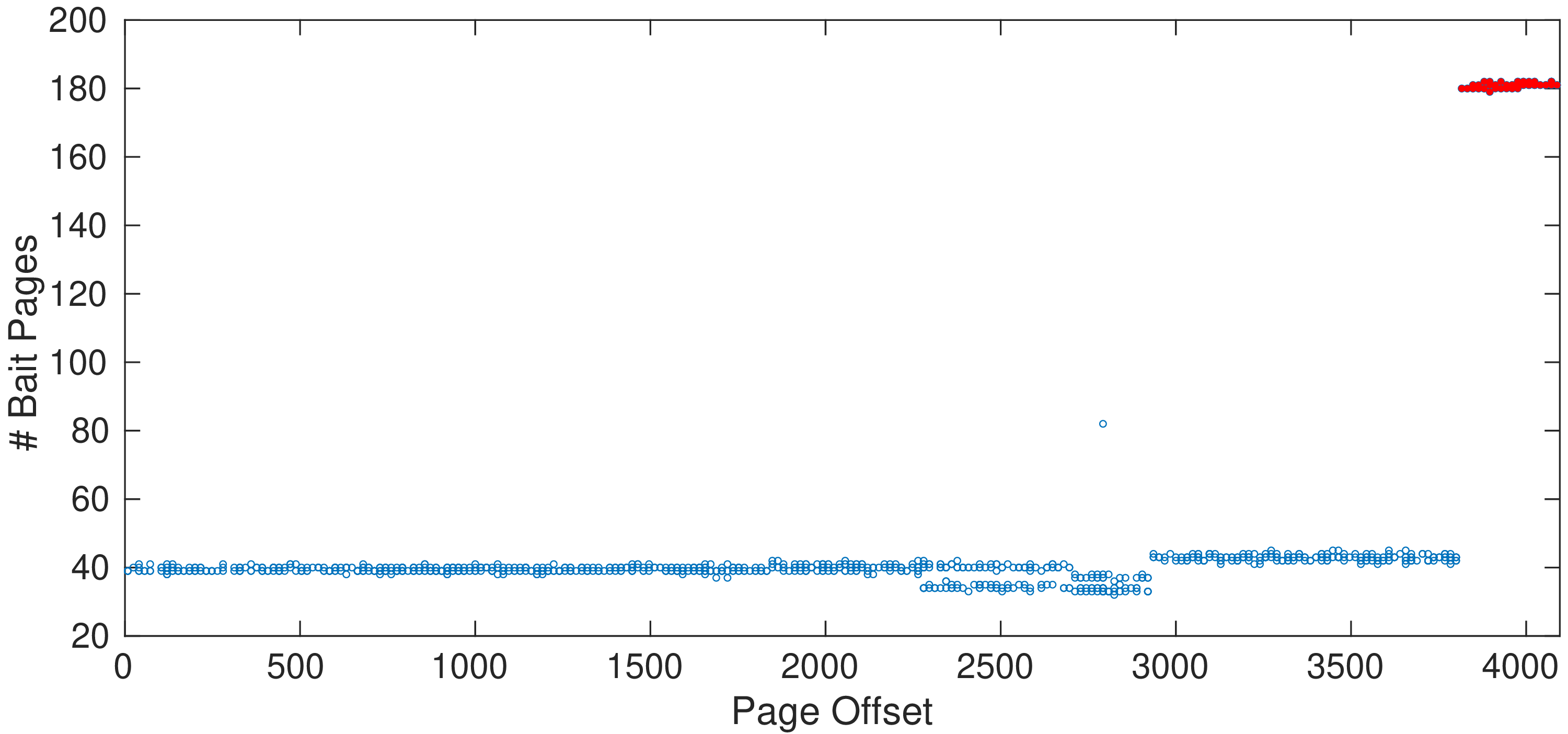}
    \caption{The relation between the number of bait pages vs. page offset of a stack variable. When the page offset is large, the number of bait pages is significantly higher (shown in red).}
    \label{fig:baits_vs_offsets}
\end{figure}

We found this relationship by unmapping pages in our attacker program and recording their physical address in a list, then in our victim program, determining where our target variable appears in the list, as well as the page offset address of the target variable (the last 12 bits). While this relationship was different for each program, it was clear that there was always a smaller set of data points where the number of bait pages clearly limited the number of possible page offsets. We created a graph of this relationship in Figure~\ref{fig:baits_vs_offsets}. We can see from the graph that if 180 bait pages were required to be released to mount the victim variable in vulnerable memory correctly, then the page offset of the said variable would be around 4000. Likewise, if the number of bait pages is 40, it can be assumed that the page offset is going to be somewhere between 0-2500. It should not be possible to find patterns in page offset because ASLR intends the offset to be based on random number generation as the offset is masked by a randomly generated value, as seen in Listing \ref{lst:aslr_linux}. 

To understand the root cause of this unusual behavior, we investigate the following methods that leak information about the page offset.

\subsection{Controlling the Page Offset with ASLR disabled}
We investigate the dependency between the number of bait pages and the page offset of a stack variable in a more controlled environment. We create the following function where a buffer with a predefined \texttt{BUFFER\_SIZE} before integer variable \texttt{var}.

\begin{figure}[!h]
\begin{lstlisting}[frame=single,
                    language=C++,numbers=none,
                    label={lst:aslr_off}
                    ]
void main(){
    char buffer[BUFFER_SIZE] = {0};
    int var = 0;
}
\end{lstlisting}
\end{figure}
\begin{figure}
    \centering
    \includegraphics[width=\columnwidth]{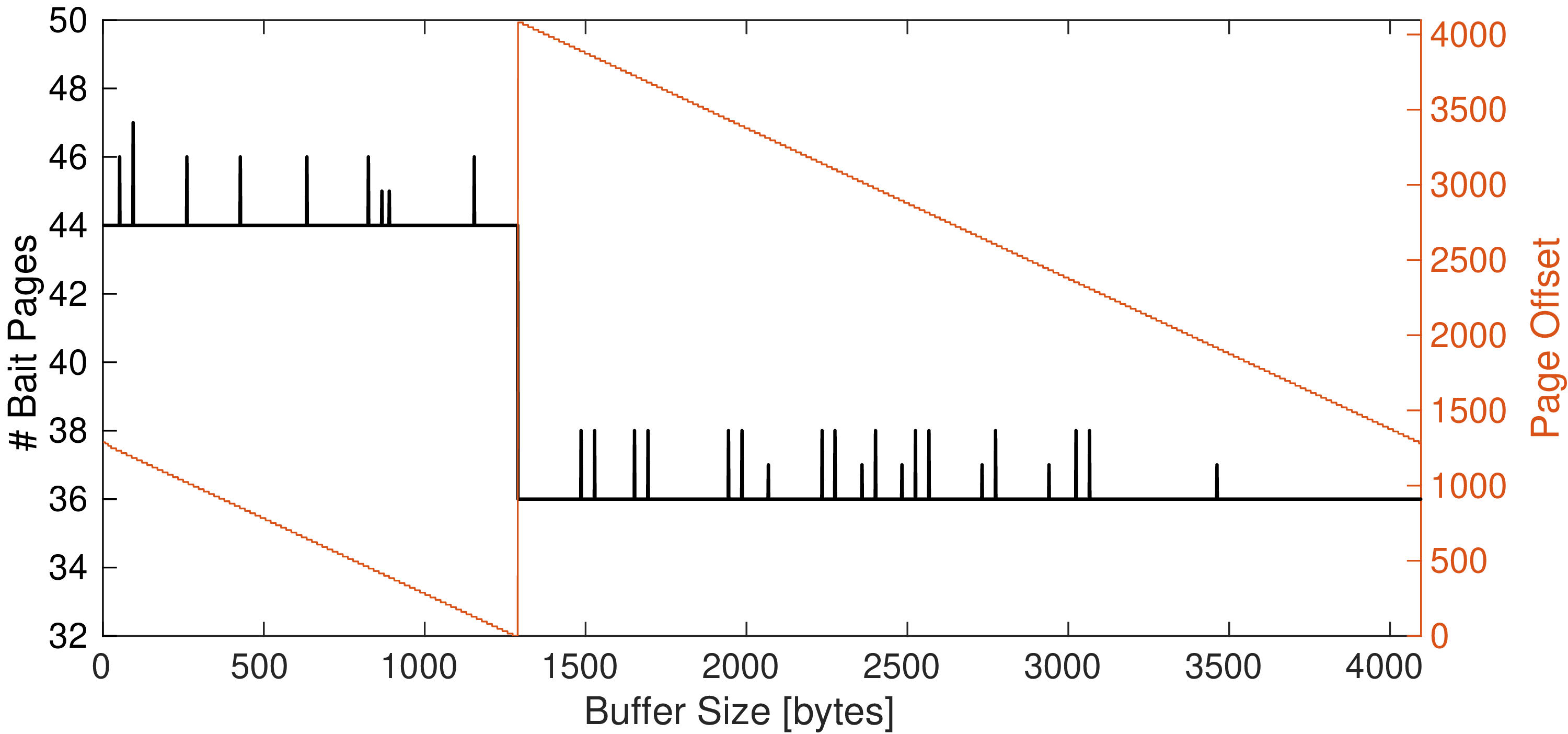}
    \caption{The dependency between the number of bait pages (black) and page offset (red) when ASLR is disabled. The page offset of the variable is manually controlled by changing the size of the buffer. The jump in the \# bait pages and page offset occurs at the same point.}
    \label{fig:aslr_off}
\end{figure}

 Note that both the buffer and the variable are stored in the stack. We disable ASLR in the system to make sure we have full control on the page offset of the variable. In Figure~\ref{fig:aslr_off}, we vary the \texttt{BUFFER\_SIZE} variable from 0 to 4K. Increasing the size of the buffer pushes the variable back in the stack and linearly decreases the page offset. We control the page offset by varying the size of the buffer. We also observe the number of required bait pages has a sudden change together with the page offset of the variable. We speculate this behavior is caused by crossing the page boundaries while increasing the \texttt{BUFFER\_SIZE} and results in an increase in the number of total pages consumed by the program.
Next, we investigate the same dependency with ASLR enabled.

\subsection{Page Fault Side Channel}

We found that monitoring for page faults gives us a side channel to determine the offset set by ASLR. A page fault will happen when a process requests data from a page in memory that is not currently loaded in DRAM. When the page fault occurs, the page needs to be moved from the \textit{swap space} in the storage to DRAM. 
There are two types of page faults; major faults and minor faults. Major faults occur when a page is requested that does not exist in memory and needs to be brought back from the swap space. A minor fault is less performance degrading and occurs when the page is currently in memory and needs to be swapped back out to the disk (usually to free up space in DRAM for other pages).

Looking at Figure \ref{fig:page_faults}, we can see that if the process receives 275 page faults (marked in red), we can guarantee that the location of the offset in DRAM is going to be somewhere between 200 and 800, which reduces the search space and randomization of the ASLR offset bits by more than a factor of 6. Additionally, if 286 page faults are detected, we know that the offset will generally \emph{not} be between 200-800, which also reduces the search space. 

We are not sure why this side channel exists, but we speculate that the randomized page offset throws page faults which we can monitor using performance monitoring by the attacker. It is important to note that this performance monitoring, e.g., the \texttt{/usr/bin/time} command, \emph{do not} require special permission to run and thus are practical to use in a real attack. 


\subsection{Remapping Pages Side Channel}

One technique we used was page remapping, where we would \texttt{unmap} $n$ pages of our attacker program, launch our victim process, then remap n pages back to our attack program. If we unmapped 500 pages, launched our victim process, then remapped 300 pages back to our attacker program, we would assume that the number of bait pages our victim required was 200 pages.
We found a slight correlation in the data, but ultimately, it was too noisy to be useful. We speculate this is because remapping pages pulls from unpredictable pools of memory, so the number of pages is not zero-sum.

\subsection{Exploiting Offset Randomization}~\label{sec:exploit_aslr}

Although ASLR is built as a security measure to prevent memory attacks, it can be exploited to make the Rowhammer attack more powerful. We propose a technique named \emph{relaunching} to exploit ASLR for Rowhammer. 

The attacker first profiles the memory to find a flippy bit location in memory. In some DRAMs, these flippy locations may be rare. For some Rowhammer enabled attacks, that require a specific bit in the page to be flippy, the attack will become less viable. In our attack, instead, we first find a flippy bit in memory, then perform the following steps:

\begin{enumerate}[noitemsep,topsep=0pt]
    \item After finding a flippy bit location, the attacker frees memory to the system containing the number of bait pages followed by the flippy page;
    \item The attacker launches the victim process which fills the recently deallocated pages;
    \item The attacker performs the Rowhammer attack on the victim process (not knowing if the flippy bit aligns with the bits required to be flipped in the victim process);
    \item The victim process ends and the attacker process \emph{remaps the memory used by the victim process back to itself and repeats the attack with the same flippy row}.

\end{enumerate}

With this approach, theoretically, the attacker only needs to find \emph{a single flippy bit in the whole system} for the attack to work. This is a dramatic improvement over other Rowhammer attacks where extensive profiling is required, and often thousands of flips are required before a successful attack. 

Relaunching works because ASLR will put the variable into a new location in the page the next time it runs. This means that rather than looking for a new flippy bit that might colocate where a flip is needed in the victim process, the victim can simply be relaunched and ASLR reshuffles the variable somewhere else, potentially into the location where it can be flipped by the flippy bits in the page.